\pdfoutput=1

\documentclass[final, 1p, times, authoryear]{elsarticle}

\usepackage{amssymb}

\usepackage{amsmath}
\usepackage{amsfonts}
\usepackage{graphicx}
\usepackage{ulem}
\def\rot{\rotatebox}
\usepackage{multicol}
\usepackage{booktabs}
\usepackage{subcaption}
\usepackage{mathrsfs}
\usepackage{lscape}
\usepackage{bm}
\usepackage[export]{adjustbox}[2011/08/13]

\journal{arXiv}

\begin{document}

\begin{frontmatter}



\title{Optimal curtailed designs for single arm phase II clinical trials}

%
%

\author[1]{Martin Law\corref{cor1}}%
\author[2]{Michael J. Grayling}
\author[1,3]{Adrian P. Mander}
\cortext[cor1]{Corresponding author}
\address[1]{Hub for Trials Methodology Research, Medical Research Council Biostatistics Unit, University of Cambridge, United Kingdom}
\address[2]{Institute of Health and Society, Newcastle University, United Kingdom}
\address[3]{College of Biomedical and Life Sciences, Cardiff University, United Kingdom}

\begin{abstract}
In single-arm phase II oncology trials, the most popular choice of design is Simon's two-stage design, which allows early stopping at one interim analysis. However, the expected trial sample size can be reduced further by allowing curtailment. Curtailment is stopping when the final go or no-go decision is certain, so-called non-stochastic curtailment, or very likely, known as stochastic curtailment.

In the context of single-arm phase II oncology trials, stochastic curtailment has previously been restricted to stopping in the second stage and/or stopping for a no-go decision only. We introduce two designs that incorporate stochastic curtailment and allow stopping after every observation, for either a go or no-go decision.
We obtain optimal stopping boundaries by searching over a range of potential conditional powers, beyond which the trial will stop for a go or no-go decision. This search is novel: firstly, the search is undertaken over a range of values unique to each possible design realisation. Secondly, these values are evaluated taking into account the possibility of early stopping. Finally, each design realisation’s operating characteristics are obtained exactly. 

The proposed designs are compared to existing designs in a real data example. They are also compared under three scenarios, both with respect to four single optimality criteria and using a loss function.

The proposed designs are superior in almost all cases. Optimising for the expected sample size under either the null or alternative hypothesis, the saving compared to the popular Simon's design ranges from 22\% to 55\%.
\end{abstract}




\end{frontmatter}



\section{Introduction}

Most novel treatments are found to be inefficacious, which makes the average development cost associated with each successful treatment extremely high \citep{trialfail}. Further, trials themselves are expensive to run \citep{trialcost}, and the nature of evaluating treatment response in oncology trials means that results are not immediately available, meaning that the trials can take time to run. This makes novel designs that can improve clinical research extremely valuable. Phase II clinical trials for cancer treatments often have a binary outcome, based on change in tumour size as measured by the RECIST scale \citep{recist}, and in phase II, these trials often contain a single arm. 

Simon's design is the most frequently used phase II design among UK clinical trials units, and alongside single-stage trials, accounts for most phase II oncology trial designs across the world \citep{jaki, simonworld}. In this design, an interim analysis takes place once a specified number of results are available. At this point, the trial stops for a no-go decision if the number of (positive) responses is not greater than some specified value, otherwise the trial continues, recruiting the remaining patients and continuing until results are available for all patients. The possibility of stopping for futility at this interim point before the end of the trial has the effect of reducing the expected sample size compared to a single-stage design \citep{simon}.

Many extensions to Simon's design have been proposed with the aim of decreasing expected sample size: for example, allowing stopping for a go decision at the interim analysis if the number of responses is greater than some specified value \citep{shuster_2stage_go}. Another extension allows stopping for either go or no-go decisions as soon as the final trial decision is certain, known as non-stochastic curtailment (NSC) \citep{chi_NSC}. The final extension we note is to allow stopping for either a go or no-go decision as soon as either decision becomes highly likely, known as stochastic curtailment (SC). This approach has been described by  \cite{jennturn} and applied to phase II clinical trials by \cite{kunz} and \cite{AR}.

We generalise the third of these extensions, presenting two designs that reduce the expected sample size: firstly, we propose a two-stage design that allows SC after each participants' results, for either a go or no-go decision. 

Secondly, we propose a single-stage design that again allows SC after each participants' results for a go or no-go decision. The methodology behind these designs, and the approaches used to compare them to existing designs, are summarised below. 

Stochastic curtailment in this paper is based on conditional power, though other approaches exist, for example, the predictive power approach and the the parameter-free approach \citep{jennturn}. The conditional power (CP) is the probability of rejecting the null hypothesis, conditional on the number of participants and responses so far, and assuming an efficacious treatment. In our proposed designs, the trial will end as soon as the CP crosses some specified upper or lower threshold. We present equations for calculating conditional power that are accurate for both stochastic and non-stochastic curtailment. Further, designs are found by searching over a range of thresholds for CP specific to each possible trial, a novel approach. 
Designs have previously been proposed that employ stochastic curtailment using a uniform sequence of conditional power thresholds, and that allow stochastic curtailment for a no-go decision only \citep{kunz, AR}. We use a real data example to compare the operating characteristics of our proposed design types to these.

The operating characteristics of curtailed designs can be estimated using simulation. However, we obtain the exact distribution for each trial's outcomes, and so each trial's operating characteristics can be found without simulation error. Further, rather than find the optimal non-curtailed design and apply stochastic curtailment to it, as in the paper of \cite{kunz}, we take curtailment into account when searching for the optimal realisation of a design. This means a wider range of possible designs can be examined. 

Designs can be compared using a number of different optimality criteria, including the expected sample size under the null ($H_0$-optimal) or alternative ($H_1$-optimal) hypotheses, and the expected sample size among the designs with the lowest maximum sample size under the null ($H_0$-minimax) or alternative ($H_1$-minimax) hypotheses. A design must also be feasible, that is, it satisfies some chosen type I error and power. A design that is optimal under a single optimality criterion is known as an optimal design. The best design often differs depending on the criterion used.

In the setting of multiple optimality criteria, \cite{jung} found optimal designs by creating a loss function that was a weighted combination of maximum sample size and expected sample size under $H_0$. Jung et al. describe a design as admissible if it has the smallest expected loss of all possible designs, for a given weighting. \cite{mander} extend this by incorporating expected sample size under $H_1$ into the loss function, and by visually examining the admissible designs in a grid covering a range of possible weights. This results in a set of admissible designs, one for each combination of weights. 
We use this approach to compare the admissible designs of each design type for a range of weights, using three sets of design parameters $(\alpha, \beta, p_0, p_1)$, termed scenarios. We then determine the admissible design for each combination of weights in each scenario. For a given combination of weights, we term the admissible design with the smallest expected loss across all tested designs the ``omni-admissible" design. We show that the omni-admissible design for almost all possible combinations of weights belongs to a proposed design, rather than an existing design.

Trials that employ early stopping are known to have bias in the na\"ive estimate of the response rate \citep{whitehead}. As such, this may be a source of concern for investigators with a strong interest in point estimation. However, there are a range of estimators available that take this bias into account for two-stage designs, and we therefore provide details of how various point estimation procedures can be extended for use with curtailed designs, and examine the resulting point estimates in terms of bias and root mean squared error \citep{porcher}.

The remainder of the paper is as follows: in the Methods section, we describe: our new designs and how they differ to currently-used designs; equations for calculating CP taking stochastic curtailment into account; the rationale for searching over a non-uniform sequence of CP thresholds; how to obtain the exact distribution of trial outcomes for a design; the loss function and some details regarding how the admissible designs are found. In the Results section, we present one real data example and three scenarios, comparing the proposed designs to existing designs for the stated optimality criteria. Generalising the optimality criteria to the loss function, we also compare designs across a grid of possible combinations of weights, and produce plots showing the best design for each combination of weights. The effect of reducing monitoring frequency is examined. For one scenario, we present plots showing the admissible designs for each design. We present plots of bias and root mean squared error in the estimates of the response rate, for a number of estimators. In the Discussion section, we summarise the results and make recommendations for investigators. The supplementary material contains more detail regarding the number of CP values there are in a trial; more detail regarding the search for admissible designs; plots showing expected loss and admissible designs for each design and scenario and more detail regarding the point estimators used.

\section{Methods}

For a trial of the considered type, the treatment is assumed to have a true response rate $p$. Let $p_0$ be the greatest response rate that is clinically uninteresting, and $p_1$ be the smallest response rate that is big enough to warrant further study. We test the null hypothesis $H_0: p=p_0$ against the alternative hypothesis $H_1: p \neq p_0$. The trial is powered under $p=p_1$.

\begin{figure}[h]
    \centering
      \includegraphics[width=\textwidth]{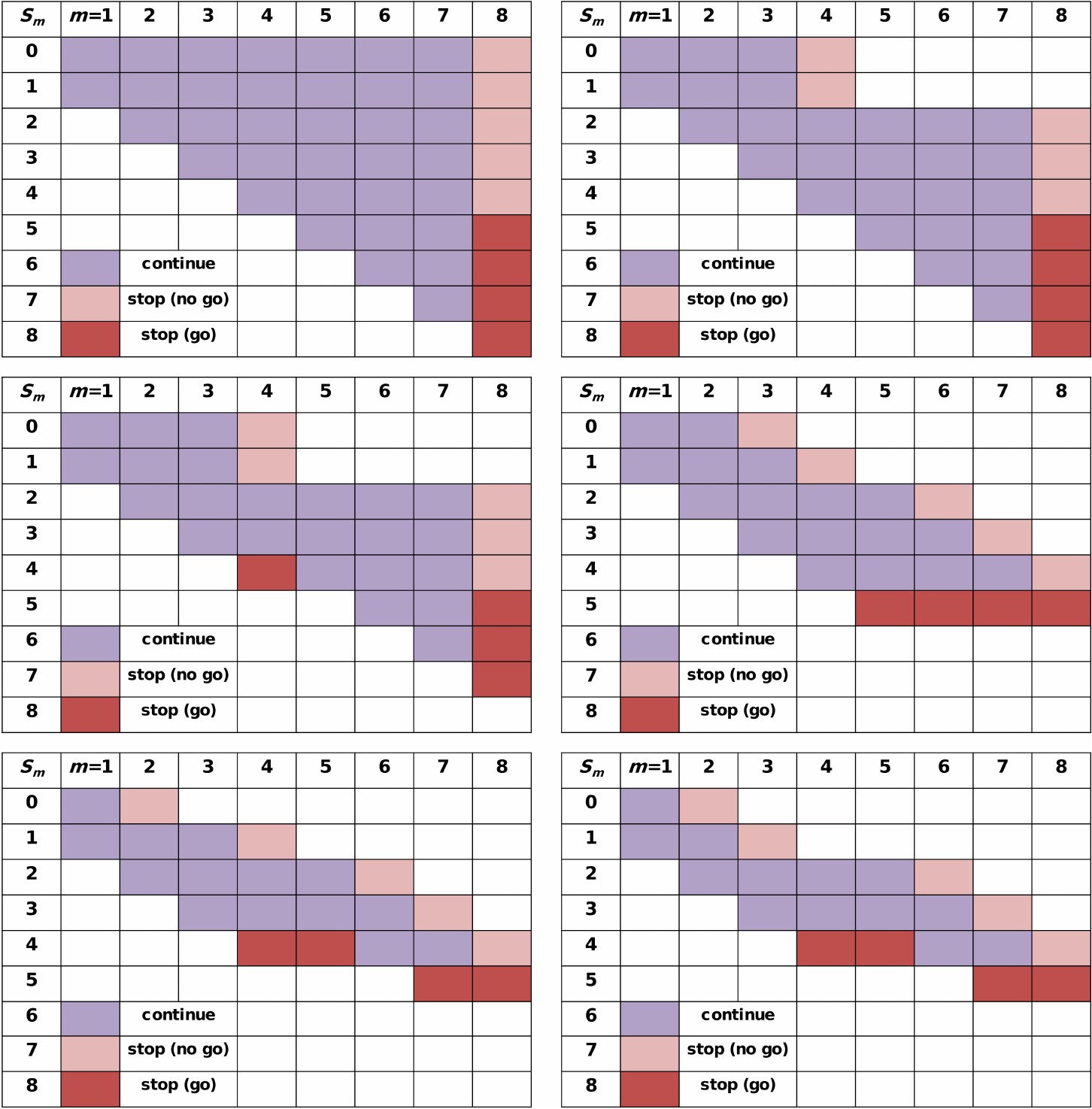}
    \caption{Illustrative diagrams of different trial designs, showing potential paths where the study would end, known as terminal points. Designs: (a) Single stage design; (b) Simon's design; (c) Simon with go; (d) NSC; (e) SC; (f) $m$-stage. $m$: Number of participant results so far. $S_m$: Number of responses so far. All trials have $N=8, r=4$, with $r_1=1$ in the two-stage designs and $e_1=3$ in Simon's design with go and no-go stopping.}\label{fig:schem}
\end{figure}

\subsection{Designs}

For single-arm trials with a binary outcome, the most simple trial design is a single-stage trial, which is comprised of $N$ patients and is deemed a success if the number of responses $S_N$ exceeds some pre-specified threshold $r$. Simon's design includes an interim analysis after $n_1$ patients, at which point the trial proceeds to the end only if the number of responses $S_{n_1}$ is greater than some $r_1$, otherwise it stops for a no-go decision \citep{simon}. \cite{shuster_2stage_go} proposed an extension to Simon's design, where the trial may stop for a go decision after $n_1$ patients if the number of responses $S_{n_1}$ exceeds some threshold $e_1$. We will refer to this design as ``Simon with go". For all designs, we use the notation $(S_m, m)$ to denote the point in a trial where $S_m$ responses have been observed after $m$ participants.

For both single-stage designs and Simon's design, it is likely that the final go or no-go decision is known before the termination of the trial: If, after $m$ patients, $S_m > r$, then the trial will be declared successful regardless of the data from the remaining patients; conversely, if $m-S_m > N-r-1$, then the trial will be declared a failure, again regardless of the data from the remaining patients. 
\cite{chi_NSC} ($CC$) proposed NSC: 
this design can be understood as Simon's design with stopping allowed as soon as the final decision is known with certainty, for either a go or no-go decision. We will refer to this as the ``NSC design".

\cite{AR} ($AR$) and \cite{kunz} ($KK$) proposed trials employing SC for a no-go decision only, i.e., stopping a trial when failure is likely but not yet certain.  A range of approaches are available for SC \citep{jennturn}, and $AR$ and $KK$ use the conditional power approach: if the probability of rejecting $H_0$, assuming $p=p_1$ and conditional on the current number of patients and responses, is below a specified threshold, $\theta_F$, the trial is ended for a no-go decision. $AR$ proposed stopping only in the second stage, while $KK$ proposed allowing stopping after each observation. $AR$ examined two thresholds, $\theta_F \in \left\lbrace 0.05, 0.10 \right\rbrace$, while $KK$ examined the uniform sequence of thresholds $\theta_F \in \left\lbrace 0, 0.01, \dots , 1 \right\rbrace$.

We propose two designs, where SC is permitted after each observation and where the trial may stop not only for a no-go decision if $CP <\theta_F$, but also for a go decision if $CP>\theta_E$, for some specified threshold $\theta_E$. Boundaries for both $\theta_E$ and $\theta_F$ may be set, for example setting $\text{min}(\theta_E)=1$ to allow SC for a no-go decision only. We have not employed such boundaries, only noting that they can readily be incorporated, as our goal is to find the optimal design realisation regardless of $\theta_E$ and $\theta_F$ values. However, boundaries may be desired, for example, for statistical reasons, or to reduce computation time. In the supplementary material, we propose and justify using the boundary $\theta_F < p_1$, and use of this boundary reduces computation time without affecting the results. The first design we propose is an extension of Simon's design and the NSC design: it can be understood as an NSC design that allows stopping if either a go or no-go decision becomes likely. This will be referred to as the ``SC design". The second design we propose can be understood as a single-stage design that allows stopping if either a go or no-go decision becomes likely. This will be referred to as the ``$m$-stage design".

The differences between the above designs are shown in Figure \ref{fig:schem}. Here, all possible points, and whether they are terminal points, i.e., points that will cause the trial to end if reached, are shown for the following designs: Single-stage, Simon's design, Simon with go, NSC design, SC design and $m$-stage design. This figure shows how these designs are related. It also shows in a practical sense how the incorporation of stochastic and non-stochastic curtailment improves expected sample size by decreasing the number of points that are possible to reach.

An objection to curtailment for a go decision is that one may wish to obtain more data if a treatment appears promising. However, the current abundance of possible treatments to be tested among relatively few participants make this argument less compelling than in the past. There may be some objections to stochastic curtailment in particular, as it allows for the termination of a trial at a point where, under another design, the final decision is not yet certain. The primary rebuttal to this is to make clear that the designs being proposed retain the desired type I error and power rates; nothing is lost by employing stochastic curtailment. A more practical argument is to note that the incorporation of stochastic curtailment into Simon's design is no different to the incorporation of the early stopping of Simon's design into a single-stage design: In the Results section, we use the example of a trial by \cite{sharma}, who used an optimal Simon design of $r_1=4, n_1=19, r=15, N=54$. Consider a single-stage trial with the same sample size and final stopping boundary, $r=15, N=54$. 
At the points in the trial where Simon's design would specify early stopping, $S_m=0, 1, 2, 3, 4$ and $m=19$, the conditional power of the single-stage trial is 0.30, 0.43, 0.56, 0.69 and 0.80 respectively. That is to say, if the trial reached the point $(S_m=4, m=19)$, the probability of a go decision would be 0.80 conditional on $p=p_1$ in a single-stage trial, but would require stopping for a no-go decision under Simon's design. Thus any acceptance of Simon's design is a tacit acceptance of ending a trial where, if a simpler design was employed, the final decision would not yet be certain.

\begin{table}[h]
\centering
\begin{tabular}{r|ccccccc}
\toprule
       			       		         			& Simon & Simon$_{GO}$ & KK & AR & NSC & SC & $m$-stage \\ 
\midrule
Allows stopping for go		        			& N	& Y & N  & N  & Y & Y & Y \\ 
Exact results (i.e., no simulation)   			& Y & Y & N  & N  & Y & Y & Y   \\ 
Allows stopping after each observation     		& N & N & Y  & N  & Y & Y & Y  \\ 
Allows NSC										& N & N & Y  & Y  & Y & Y & Y   \\
Allows SC 						      			& N & N & Y  & Y  & N & Y & Y  \\
Allows SC for go				    			& N & N & N  & N  & N & Y & Y  \\
Trial-specific possible $\theta$'s investigated & ---& ---& N  & N & --- & Y & Y  \\ 
\bottomrule
\end{tabular} 
\caption{Comparison of methods. KK=Kunz and Kieser; AR=Ayanlowo and Redden.}
\label{full_comparison}
\end{table}

\begin{table}[h]
\centering
\begin{tabular}{cccc|l}
\toprule
\rot{90}{Allows stopping for go and no-go} & \rot{90}{Allows stopping at any point} & \rot{90}{Allows NSC} & \rot{90}{Allows SC} & Two-stage design\\      			
\midrule
					 N	&N	&N	&N	& Simon  \\
					 N	&N	&N	&Y	& NA 	\\
					 N	&N	&Y	&N	& --- 	\\
					 N	&N	&Y	&Y	& Simon + AR*$\dagger$	\\
					 N	&Y	&N	&N	& NA 	\\
					 N	&Y	&N	&Y	& NA 	\\
					 N	&Y	&Y	&N	& --- 	\\
					 N	&Y	&Y	&Y	& Simon + KK**$\dagger$ 	\\
					 Y	&N	&N	&N	& Simon with go 	\\
					 Y	&N	&N	&Y	& NA	\\
					 Y	&N	&Y	&N	& --- 	\\
					 Y	&N	&Y	&Y	& NSC+AR*$\dagger$ 	\\
					 Y	&Y	&N	&N	& NA 	\\
					 Y	&Y	&N	&Y	& NA 	\\
					 Y	&Y	&Y	&N	& NSC 	\\
					 Y	&Y	&Y	&Y	& SC, NSC+KK**$\dagger$  	\\
\bottomrule
\end{tabular} 
\caption{Taxonomy of two-stage methods. KK=Kunz and Kieser; AR=Ayanlowo and Redden. *Stochastic curtailment uses $\theta_F \in \left\lbrace 0.05, 0.10\right\rbrace$. **Stochastic curtailment uses $\theta_F \in \left\lbrace 0, 0.01, \dots, 1 \right\rbrace$. $\dagger$ Stochastic curtailment for futility only. NA: Design not possible. Dash: Design possible. }
\label{tab:two_stage_comparison}
\end{table}

The differences between various existing designs and our proposed designs are also shown in Table \ref{full_comparison}, and in a taxonomy of possible two-stage designs in Table \ref{tab:two_stage_comparison}. The SC design (and the $m$-stage design) allow stopping at any point in the trial, stopping when the final go or no-go decision is certain (or likely), and exact distributions are used to obtain operating characteristics that are free from simulation error. In addition, the proposed designs allow SC for a go decision, and take curtailment into account while searching for possible designs.

\subsection{Obtaining exact distributions}

\begin{figure}[h]
 \centering
 \includegraphics[width=0.5\textwidth]{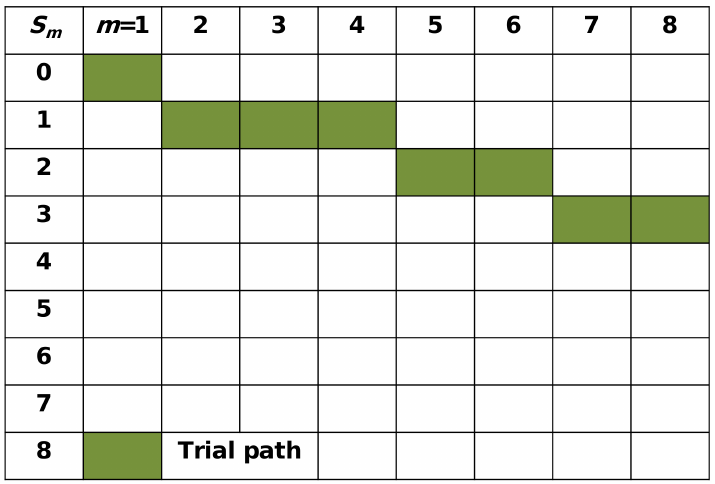}
 \caption{Example of a path for a single-stage trial with $N=8$.}
 \label{fig:path}
 \end{figure}

 There are a number of possible sequences of participant results that lead to a point $(S_m, m)$; consider each possible realisation of a trial as the ``path" of a trial. As an example, see Figure \ref{fig:path}: This example shows a possible path of a single-stage trial with no early stopping. This path may be described as $\boldsymbol{S_m}=0, 1, 1, 1, 2, 2, 3, 3, \boldsymbol{m}=1, \dots , N$. A path may end at some point $m<N$ if early stopping is permitted. The set $P_{S_m, m}$ is the set of paths that have the terminal point $(S_m, m)$.

The trials we describe progress theoretically, and often practically, one participant result at a time, where each result is response or no response. For a single-stage trial where no early stopping is allowed, the number of possible paths is $2^N$. Recording the probabilities of all possible paths would allow the exact distribution of the trial outcomes to be known. Even for a trial of moderate size, say $N=30$, the number of paths $(2^{30} > 10^9)$ would make this computationally intensive. However, to obtain the exact distribution of a trial, the probability of each path is not necessary; for any type of design, all that is required is the probability of reaching each point $(S_m, m)$ in the trial that results in a decision to stop, i.e., each terminal point. With these probabilities known, the probabilities of reaching the terminal points that result in a go decision can be summed under the assumptions $p=p_0$ and $p=p_1$ to obtain type I error and power, respectively. Moreover, it is possible to obtain the CP of a trial at any point $(S_m, m)$, as detailed below. The expected sample size can be obtained by multiplying the number of participants $m$ at each terminal point by the probability of its occurrence.  Using the exact distribution means that the operating characteristics of the trial outcomes --- for example, overall type I error, power, expected sample size and so on --- are known without recourse to simulation. This is essential when searching for the true optimal design, as simulation error could result in a design with a non-optimal expected sample size being chosen. 

\subsection{Conditional Power}

The conditional probability at point $(S_m, m)$ is the probability of rejecting $H_0$ conditional on being at point $(S_m, m)$ and assuming a true response rate $p$. Setting $p=p_1$ gives the conditional power $CP(p_1)_{S_m, m}$. 
However, from here we refer only to conditional power rather than conditional probability. 
\cite{kunz} previously provided an equation for calculating CP, but this equation did not account for early stopping. For the NSC design, we have derived the following equation for calculating $CP(p_1)_{S_m, m}$:
 
\begin{equation}\label{eq:cp}
  CP(p_1)_{S_m, m} = \left\{
  \begin{array}{@{}ll@{}}
    0, & \text{if } m-S_m > N-r-1 \\
    & \text{ or } (m-S_m > n_1-r_1-1 \text{ and } m \leq n_1) \\[10pt]
           \displaystyle\sum\limits_{j=r-S_m}^{n_1-m-1} \displaystyle \left[ A(j, r_1)  \displaystyle\sum\limits_{i=r-S_m}^{N-(j+m+1)-1} A(i,r) \right], & \text{if } m-S_m \leq n_1-r_1-1 \text{ and } m\leq n_1\\[20pt]
      \displaystyle\sum\limits_{i=r-S_m}^{N-m-1} A(i,r)& \text{if } m-S_m \leq N-r-1 \text{ and } m > n_1\\[20pt] 
    1, & \text{if } S_m > r
  \end{array}\right\}
\end{equation} 

where
\[
A(x,y) = \binom{x}{y-S_m} p_1^{y-S_m+1}(1-p_1)^{x-(y-S_m)}.
\]

The CP for the NSC design can also be written recursively as 

\begin{equation}\label{eq:cp_recursive}
  CP(p_1)_{S_m, m} = \left\{
  \begin{array}{@{}ll@{}}
    0, & \text{if } m-S_m > N-r-1 \text{ or } (m-S_m > n_1-r_1-1 \text{ and } m \leq n_1) \\[10pt]
    D, & \text{if } m-S_m \leq N-r-1  \text{ or } (m-S_m \leq n_1-r_1-1 \text{ and } m \leq n_1)\\[10pt]
    1, & \text{if } S_m > r
  \end{array}\right\},
\end{equation} 

where $D =  p_1 CP(p_1)_{S_m+1, m+1} + (1-p_1) CP(p_1)_{S_m, m+1}$.
For a single-stage trial incorporating NSC, the CP can be obtained using these equations by omitting the conditions relating to $r_1$ and $n_1$. In its recursive form, it can be seen that the CP at any point in a trial is a function of the CP at points with more responses and more participants, if such points exist. Consider a matrix of CP values for an NSC design based on number of responses $S_m$ as rows and number of participants $m$ as columns. As an example, let the maximum sample size be $N=8$ and the stopping boundary be $r=4$, as in Figure \ref{fig:schem} (d). The CP at the point ($S_m=3, m=4$), $CP(p_1)_{3,4}$, is a function of $CP(p_1)_{3,5}$ and $CP(p_1)_{4,5}$, which in turn are a function of $CP(p_1)_{3,6}$ and $CP(p_1)_{4,6}$, and $CP(p_1)_{4,6}$ and $CP(p_1)_{5,6}$ respectively. 

For designs that incorporate NSC, the trial stops and a no-go decision is taken if $CP=0$. The trial stops and a go decision is taken if $CP=1$. The $CP$ at any point can be obtained using Equation (\ref{eq:cp}) directly. For designs that incorporate SC, the trial will additionally end at any point where $0 < CP<\theta_F$ or $\theta_E < CP < 1$. As the CP is a function of later points in the trial, the predetermined decision to end a trial at any point where $0 < CP<\theta_F$ effectively causes the CP of such points to be rounded to zero. Conversely, points where $\theta_E < CP < 1$ are rounded to one. This in turn affects the CP of earlier points in trial. As such, when incorporating SC, it is logical to calculate CP at each point using a recursive equation, one value at a time, starting at the point $(S_m=r, m=N-1)$, where $CP(p_1)_{r, N-1}=p_1$. All ``earlier" points, in the trial, i.e., points such that $m<N-1$, are either a function of $CP(p_1)_{r, N-1}$ or are stopping boundaries. For points with more responses or more participants, $CP(p_1)_{a, N}=0$ if $a \leq r$, and $CP(p_1)_{a, b}=1$ for any $a > r$ and any $b$.  
Thus for the SC design, the CP at each point can be obtained using the following equation:

\begin{equation}\label{eq:cp_sc}
  CP(p_1)_{S_m, m} = \left\{
  \begin{array}{@{}ll@{}}
    0, & \text{if } D < \theta_F \text{ or }  m-S_m > N-r-1 \text{ or } (m-S_m > n_1-r_1-1 \text{ and } m \leq n_1)  \\[10pt]
    D, & \text{if }  \theta_F \leq D \leq \theta_E \text{ and } \{ m-S_m > N-r-1 \text{ or } (m-S_m > n_1-r_1-1 \text{ and } m \leq n_1) \} \\[10pt]
    1, & \text{if } D > \theta_E  \text{ or } S_m > r
  \end{array}\right\}.
\end{equation} 

As an analogue to the equations for NSC designs, Equation (\ref{eq:cp_sc}) can also be used to obtain the CP for the $m$-stage design, which is essentially a single-stage design that incorporates SC, by omitting the conditions relating to $r_1$ and $n_1$.

\subsection{Choosing thresholds $\theta_F$ and $\theta_E$}

\begin{figure}[h]
\centering
\includegraphics[width=0.6\textwidth]{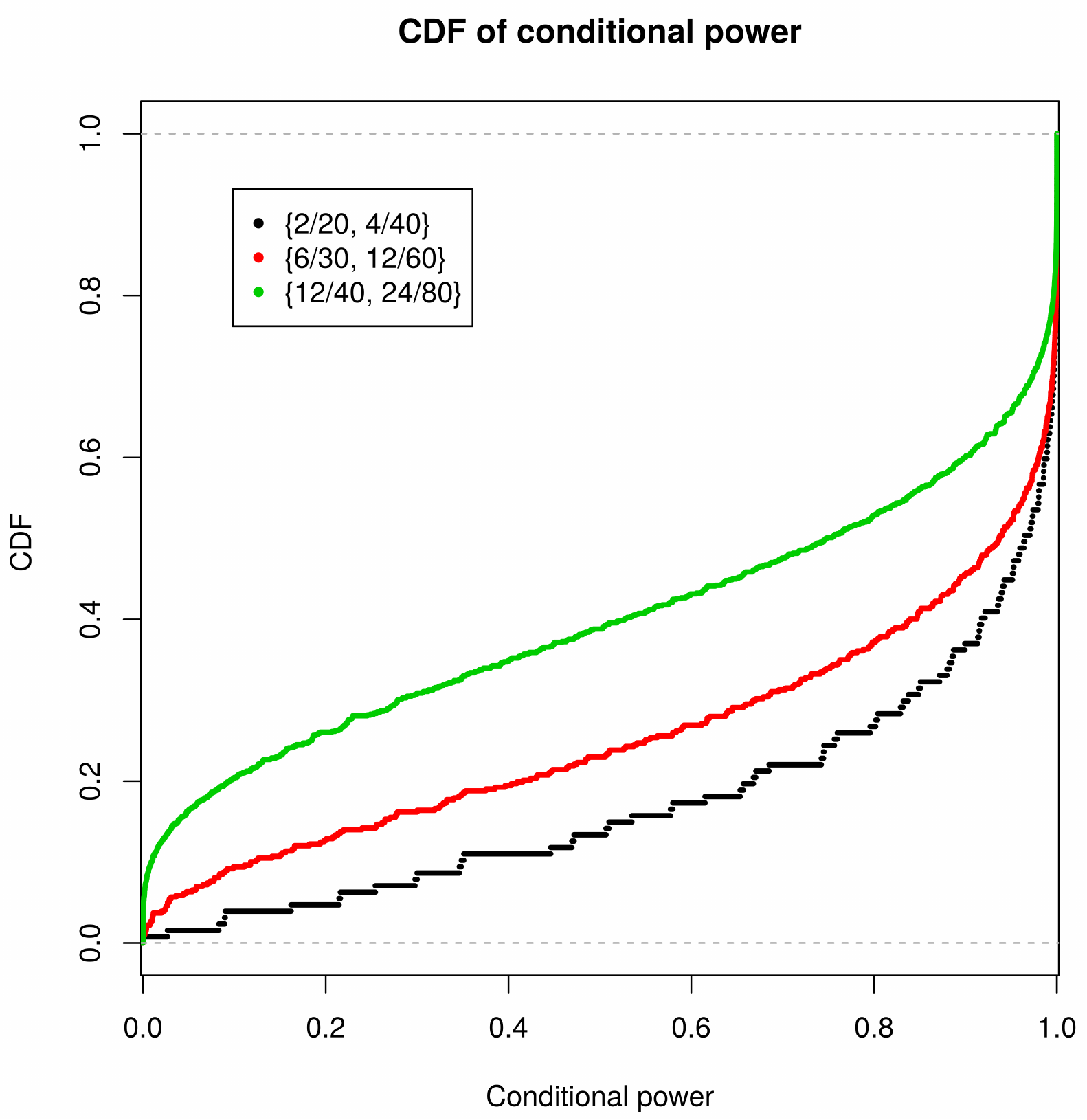}
\caption{Cumulative distribution function of unique conditional power values for NSC designs  (format $\{r_1/n_1, r/N\}$) with $r=Np_0, N(p_0+p_1)/2$ and $Np_1$ for $N=40, 60$ and $ 80$ respectively, $r_1=r/2$, $n_1=N/2$, $p_0=0.1$ and $p_1=0.3$.}
\label{fig:CP_distn}
\end{figure}

Define $\boldsymbol{\theta}$ as the set of values from which ordered pairs of $\theta_F$ and $\theta_E$ (such that $\theta_F < \theta_E$) will be created and searched over to find admissible designs. 
One could use a uniformly distributed set of possible thresholds to some specified degree of coarseness. However, by choosing a uniform, coarse set of values of $\boldsymbol{\theta}$ to search over, some designs with good operating characteristics may be missed. Conversely, undertaking a fine search would be computationally intensive --- for example, the sequence $\left\lbrace 0, 0.001, \dots, 1 \right\rbrace$ produces 500,500 ordered pairs $(\theta_F, \theta_E)$. Aside from the computational intensity required, undertaking a search over a fine uniform grid may still result in missing potential designs. This is because the effect of a chosen threshold $\theta_F$ or $\theta_E$ depends on the conditional power at each possible point in the trial. Consider for example, the NSC design shown in Figure \ref{fig:schem} (d): for each possible combination of participants so far, $m$, and number of responses, $S_m$, there exists some conditional power, $CP(p_1)_{S_m, m}$. At the points where the trial stops for a go or no-go decision, the CP is equal to one or zero respectively, and is strictly between these values at all other points. Further, in trials of this type, the CP values are not uniformly distributed; instead, most of the mass is close to zero or one. This is shown for three example trials in Figure \ref{fig:CP_distn}. To account for the lack of a uniform distribution, we propose searching over a set of thresholds chosen based on the conditional power at each point in each possible trial. We note every possible value of $CP(p_1)_{S_m, m}$, including zero and one, for a given combination of $\lbrace r, N \rbrace$ or $\lbrace r_1, n_1, r, N \rbrace$, and use this as the trial-specific set $\boldsymbol{\theta}$. We search over ordered pairs $(\theta_F, \theta_E)$ from elements of this set such that $\theta_F < \theta_E$. For an NSC design with typical design parameters $(r_1=5, n_1=20, r=10, n=40, p_0=0.1, p_1=0.3)$, there are 239 possible CP values, from which 28,441 possible ordered pairs can be created. Placing sensible constraints on the ordered pairs, such as $\theta_F < p_1, \theta_E \geq 0.95$, reduces the number of possible ordered pairs to 7,809, with such constraints justified in the supplementary material.

\subsection{Admissible design search}

For the proposed designs, possible design realisations are found by first setting the desired type I error and power, $p_0$ and $p_1$, and a range for $N$. For each $N$, a sensible range for $r$ is chosen based on bounds suggested by either \cite{ahern} or \cite{wald}; see the supplementary material for details. The combinations of $\lbrace r, N \rbrace $ are stored, as are possible combinations of $r_1$ and $n_1$ in the case of two-stage designs. Let $\mathscr{R}$ be the set of all possible combinations of $\lbrace r, N \rbrace $ or $\lbrace r_1, n_1, r, N \rbrace $. For each $\mathscr{R}$, the CP of each point in that trial is obtained. These CPs form the trial-specific set $\boldsymbol{\theta}$. Within each $\mathscr{R}$, type I error, power, expected sample size under $p=p_0 (E(N|p=p_0))$ and $p=p_1 (E(N|p=p_1))$ are found for ordered pairs $( \theta_F, \theta_E ) \in \boldsymbol{\theta}: \theta_F < \theta_E$. The dominated designs --- designs that are not superior to at least one other design in terms of $E(N|p_0)$, $E(N|p_1)$ or $N$ --- are discarded. What remains is a collection of admissible designs. The expected sample sizes of the $H_0$- and $H_1$-optimal and  $H_0$- and $H_1$-minimax admissible designs of the proposed designs will be compared to those of Simon's design, Simon with go and the NSC design, and additionally to those of the designs found using the sequential probability ratio test of \cite{wald} in the case of the  $H_0$- and $H_1$-optimal criteria. Wald's sequential probability ratio test is a design that produces trials that are feasible and have low  $E(N|p_0)$ and $E(N|p_1)$, but no maximum sample size. More details of this design  are provided in the supplementary material.

Note that type I error and power are not calculated prior to adding curtailment, as a design that is feasible under SC may not be feasible before the incorporation of SC, and so if such designs were discarded in advance, they would be missed. An example of such an instance is provided in the supplementary material.

\subsection{The loss function}

\cite{jung} introduced the concept of choosing a design based not on a single optimality criterion, but instead on a combination of two optimality criteria, weighted in importance by an investigator. This was extended by \cite{mander} to an expected loss function with weights on three optimality criteria: Expected sample size under the null hypothesis $E(N|p_0)$, the expected sample size under the alternative hypothesis $E(N|p_1)$ and maximum sample size $N$. The expected loss function is

\[ L = q_0 E(N|p_0) + q_1 E(N|p_1) + (1-q_0-q_1)N, \]

where $q_0, q_1 \in [0, 1] \text{ and } q_0 + q_1 \leq 1$. In Mander et al., the admissible designs --- defined as the design realisations with the smallest expected loss for a given set of weights --- were plotted for a grid of all possible combinations of weights. We extend this concept to allow the comparison of design realisations across differing designs: for each combination of weights, the realisation with the lowest loss, $L$, across all designs is found, and the design to which it belongs is termed the ``omni-admissible" design for that combination of weights. Across a grid of possible combinations of weights, the design to which each omni-admissible design belongs is plotted. In addition, the difference between the expected loss of two designs at each set of weights is quantified, for certain pairs of designs. The values have no intrinsic meaning; their only purpose is to facilitate the comparison of designs. For brevity, the set of admissible designs for each design is plotted for the first scenario only, with the remainder in the supplementary material. From these plots, the number of admissible designs, and the range of weights for which each admissible design has the lowest loss, can be seen. 

\subsection{Inference: Estimation of response rate}
Though not the main aim of phase II trials, it is of importance to undertake inference using the final trial data in order to help make decisions about possible future trials. In particular, one may estimate the response rate of a treatment. The MLE of the response rate is the observed response rate $\hat{p} = S_m/m$, where $m$ represents the number of participants after which the trial stopped. However, this estimator is biased in trials that allow stopping at an interim analysis, and as such there are a number of estimators available with the aim of reducing this bias \citep{porcher}. However, such estimators have only been previously presented for two-stage designs. To address possible concerns regarding point estimation in curtailed designs, we examine estimates of the response rate across existing and novel designs, from five estimators, extended to the multi-stage case: the na\"ive estimator, that is, the MLE above; the bias-adjusted estimator \citep{bias_adj_chang}; the simplified bias subtraction estimator \citep{bias_sub_guo}; the median unbiased estimator (MUE) \citep{mue} and the uniformly minimum-variance unbiased estimator (UMVUE) \citep{umvue}. A range of estimators are used as there is no single estimator that performs best in all situations. We report the bias, $Bias(\hat{p}|p) = E(\hat{p}|p) - p$, and the RMSE. The RMSE, $RMSE(\hat{p}|p) = \sqrt{(Bias(\hat{p}|p))^2 + Var(\hat{p}|p)}$, where $Var(\hat{p}|p) = E(\hat{p}^2|p) - E(\hat{p}|p)^2$, is equivalent to taking the square root of the weighted average of the squared distances between each possible point estimate and the true value. For brevity, more details regarding each estimator is provided in the supplementary material. The results are shown for the $H_0$-optimal admissible designs of each design. 



\section{Results}

\subsection{Real data example}

\cite{kunz} present a real data example from \cite{sharma}. In this trial, the following design parameters were chosen: $ \alpha=0.05, \beta=0.1, p_0=0.2, p_1=0.4$. Kunz and Kieser compare the following combinations of designs to the optimal Simon design: optimal Simon design with NSC for no-go only, SC for no-go only, and early stopping in stage 2 only (``Simon + AR"); optimal Simon design with NSC for no-go only, SC for no-go only, and early stopping in both stages (``Simon + KK"); optimal Simon design with NSC for go and no-go, SC for no-go only, and early stopping in stage 2 only (``CC + AR"); optimal Simon design with NSC for go and no-go, SC for no-go only, and early stopping in both stages (``CC + KK"). The threshold $\theta_F=0.4$ is used, as Kunz and Kieser directly report the results for only $\theta_F=0.4$ and $\theta_F=0.6$, stating that trials using $\theta_F=0.6$ do not achieve requisite power. Table \ref{tab:kk_comparison} contains the operating characteristics for these designs to as great an extent as possible, in addition to those from: Simon's design; the NSC design (``CC"); two SC designs, the former chosen for its resemblance to the other compared trials in terms of maximum sample size $N$, the latter chosen as it is the $H_0$-optimal design; two $m$-stage designs, chosen in the same manner as the SC designs, and finally Wald's sequential probability ratio test \citep{wald}. The operating characteristics of Simon+AR, Simon+KK, CC+AR and CC+KK were obtained from the results of \cite{kunz} and from Stata using the \textit{simontwostage} package \citep{simontwostage}. The maximum $N$ searched over for the SC and $m$-stage designs respectively is $N=58$ and $N=94$, due to computational intensity, with the range of $r$ chosen based on the boundaries of \cite{wald} (see supplementary material).

It can be seen from Table \ref{tab:kk_comparison} that with the exception of Wald, the designs with the lowest $E(N|p_0)$ are Simon+KK and CC+KK, which use a threshold of $\theta_F=0.4$ and allow stopping at any point. However, these designs did not obtain the requisite type I error of 0.05 in the simulation undertaken. The four trials from the two proposed designs achieve a lower  $E(N|p_0)$ than all feasible designs, with the exception of Wald, while maintaining the necessary type I and type II errors. They also have lower thresholds for stopping for a no-go decision compared to other designs that use SC, with a maximum of $\theta_F=0.228$ compared to $\theta_F=0.4$. Further, the first $m$-stage design has a lower maximum sample size than all other designs.

The study ended at the first stage, with 1 response out of 19 participants. Using NSC only, the study would have ended after 16 participants. However, using the $m$-stage design optimised for $E(N|p_0)$, the study would have ended after 8 to 11 participants, the final stopping point dependent on when the single response occurred.

\begin{table}[h]
\centering
\begin{tabular}{cccccccccccc}
\toprule
Design & $r_1$ & $n_1$ & $r$ & $n_2$ & $N$ & $\alpha$ & Power & $E(N|p_0)$ & $E(N|p_1)$ & $\theta_F$ & $\theta_E$ \\ 
\midrule
Simon 		& 4 & 19 & 15 & 35 & 54 & 0.048 & 0.904 & 30.4 & 51.6 & --   & --\\ 
CC 			& 4 & 19 & 15 & 35 & 54 & 0.048 & 0.904 & 28.2 & 37.6 & 0.000 & 1.000\\ 
Simon + AR 	& 4 & 19 & 15 & 35 & 54 & -- & 0.90 & 26.6 & -- & 0.400 & 1.000 \\ 
Simon + KK 	& 4 & 19 & 15 & 35 & 54 & 0.054 & 0.904 & 21.3 & -- & 0.400 & 1.000 \\ 
CC + AR 		& 4 & 19 & 15 & 35 & 54 & -- & 0.90 & 25.4 & -- & 0.400 & 1.000 \\ 
CC + KK 		& 4 & 19 & 15 & 35 & 54 & 0.054 & 0.904 & 21.0 & -- & 0.400 & 1.000 \\
\hline
SC$_1$		& 2 & 14 & 15 & 40 & 54 & 0.050 & 0.901 & 23.0 & 26.6 & 0.164 & 0.998	\\
SC$_2$		& 4 & 21 & 17 & 37 & 58 & 0.050 & 0.900 & 22.6 & 25.5 & 0.199 & 0.998	\\
$m$-stage$_1$& --& -- & 15 & -- & 52 & 0.049 & 0.909 & 25.3 & 25.8 & 0.135 & 0.996 \\ 
$m$-stage$_2$& --& -- & 26 & -- & 94 & 0.049 & 0.902 & 22.1 & 23.3 & 0.228 & 0.998 \\ 
\hline
Wald			& --& -- & -- & -- & -- & 0.050 & 0.900 & 21.8 & 22.7 & -- & -- \\ 
\bottomrule
\end{tabular} 
\caption{Comparison of designs, with design parameters $(\alpha, \beta, p_0, p_1)=(0.05, 0.10, 0.20, 0.40)$. $CC$ identical to NSC design. Blanks in $\alpha$ and $E(N|p_1)$ due to data not being included in Kunz and Kieser and not being reproducible.}
\label{tab:kk_comparison}
\end{table}

\subsection{Example trials}
Three sets of design parameters, or scenarios, were used to compare five designs: Simon's design; Simon with go; the NSC design; the SC design and the $m$-stage design. For each scenario and design, a set of admissible designs was obtained as described above. For Simon and Simon with go, the maximum sample size searched over was 20$\%$ greater than the maximum sample size of the $H_0$-optimal design, as in \cite{mander}. For NSC and $m$-stage, the maximum sample size searched over was set to 80, approximately 2-3 times greater than the maximum sample size for the admissible Simon's designs under $H_0$-optimal and $H_0$-minimax. For SC, maximum sample size was 43 to 47, due to computational intensity. For the proposed designs, the range of $r$ was chosen based on the bounds of \cite{ahern} $\{\left \lfloor{Np_0}\right \rfloor, \left \lceil{Np_1}\right \rceil\}$, though the final set of admissible designs all contained values of $r$ that were also within the (stricter) bounds of \cite{wald}. Also reported is $E(N|p_0)$ and $E(N|p_1)$ from Wald's sequential probability ratio test. As Wald's probability ratio test seeks to minimise expected sample size and has no maximum sample size, the expected sample sizes from this test will be compared to those obtained under the $H_0$- and $H_1$-optimal criteria.

\subsection{Scenario 1: Design parameters $(\alpha, \beta, p_0, p_1)=(0.05, 0.15, 0.1, 0.3)$}
\begin{table}[ht]
\centering
\begin{tabular}{rcccccccccccc}
  \toprule
 & $r_1$ & $e_1$ & $n_1$ & $r$ & $n_2$ & $n$ & $E(N|p_0)$ & \%S$_{p_0}$ & $E(N|p_1)$ & \%S$_{p_1}$ & $\theta_F$ & $\theta_E$ \\ 
  \midrule $\bm{H_0}$\textbf{-optimal} \\ 
Simon & 1 &  & 11 & 6 & 24 & 35 & 18.3 & 1.00 & 32.3 & 1.00 &  &  \\ 
  Simon go & 1 & 4 & 11 & 6 & 24 & 35 & 18.2 & 1.00 & 27.2 & 0.84 &  &  \\ 
  NSC & 1 &  & 13 & 5 & 15 & 28 & 17.6 & 0.97 & 18.5 & 0.57 & 0.000 & 1.000 \\ 
  SC & 4 &  & 27 & 7 & 14 & 41 & 14.3 & 0.78 & 15.0 & 0.46 & 0.186 & 0.993 \\ 
  $m$-stage &  &  &  & 13 &  & 80 & 14.1 & 0.77 & 14.4 & 0.45 & 0.226 & 0.997 \\ 
   \midrule $\bm{H_1}$\textbf{-optimal} \\Simon & 2 &  & 18 & 5 & 9 & 27 & 20.4 & 1.00 & 26.5 & 1.00 &  &  \\ 
  Simon go & 0 & 3 & 13 & 6 & 17 & 30 & 25.1 & 1.23 & 20.0 & 0.76 &  &  \\ 
  NSC & 1 &  & 13 & 5 & 15 & 28 & 17.6 & 0.87 & 18.5 & 0.70 & 0.000 & 1.000 \\ 
  SC & 4 &  & 24 & 8 & 19 & 43 & 15.5 & 0.76 & 14.6 & 0.55 & 0.126 & 0.984 \\ 
  $m$-stage &  &  &  & 12 &  & 66 & 14.3 & 0.70 & 14.4 & 0.54 & 0.189 & 0.990 \\ 
   \midrule \\Wald &  &  &  &  &  &  & 13.9 & 0.68 & 13.9 & 0.52 &  &  \\ 
   \midrule $\bm{H_0}$\textbf{-minimax} \\Simon & 2 &  & 18 & 5 & 9 & 27 & 20.4 & 1.00 & 26.5 & 1.00 &  &  \\ 
  Simon go & 1 & 4 & 14 & 5 & 13 & 27 & 19.3 & 0.95 & 21.0 & 0.79 &  &  \\ 
  NSC & 2 &  & 18 & 5 & 9 & 27 & 19.3 & 0.95 & 18.7 & 0.71 & 0.000 & 1.000 \\ 
  SC & 0 &  & 10 & 5 & 17 & 27 & 17.1 & 0.84 & 16.3 & 0.62 & 0.070 & 0.990 \\ 
  $m$-stage &  &  &  & 5 &  & 27 & 18.7 & 0.92 & 16.6 & 0.63 & 0.084 & 0.990 \\ 
   \midrule $\bm{H_1}$\textbf{-minimax}  \\Simon & 2 &  & 18 & 5 & 9 & 27 & 20.4 & 1.00 & 26.5 & 1.00 &  &  \\ 
  Simon go & 1 & 4 & 15 & 5 & 12 & 27 & 20.3 & 0.99 & 20.8 & 0.78 &  &  \\ 
  NSC & 2 &  & 18 & 5 & 9 & 27 & 19.3 & 0.95 & 18.7 & 0.71 & 0.000 & 1.000 \\ 
  SC & 4 &  & 24 & 5 & 3 & 27 & 18.8 & 0.92 & 15.8 & 0.60 & 0.050 & 0.986 \\ 
  $m$-stage &  &  &  & 5 &  & 27 & 18.7 & 0.92 & 16.6 & 0.63 & 0.084 & 0.990 \\ 
   \bottomrule
\end{tabular}
\caption{Admissible designs for each design type, Scenario 1: $(\alpha, \beta, p_0, p_1)=(0.05, 0.15, 0.10, 0.30)$. For all designs, requisite type I error and power is reached. Columns \%S$_{p_0}$ and \%S$_{p_1}$ show expected sample size as a proportion of Simon's design under $p=p_0$ and $p=p_1$ respectively.} 
\label{tab:scenario1}
\end{table}

Table \ref{tab:scenario1} shows the optimal design for each compared design, for four optimality criteria: $H_0$-optimal, $H_1$-optimal, $H_0$-minimax and $H_1$-minimax. For all four optimality criteria, the optimal designs of the proposed designs outperform those of the existing two-stage designs, and employ thresholds of $\theta_F < 0.23$ and $\theta_E > 0.98$ in each case. With the exception of the SC design under $H_1$-optimal, the proposed designs have lower expected sample sizes under both the null and alternative hypotheses. The expected sample sizes from Wald's test are $E(N|p_0)=13.9, E(N|p_1)=13.9$, comparable to those of the $m$-stage design. Note that under the minimax criteria, the SC design happens to be the $m$-stage design with the addition of one explicit interim analysis.

\begin{figure}
    \centering
        \includegraphics[width=\textwidth]{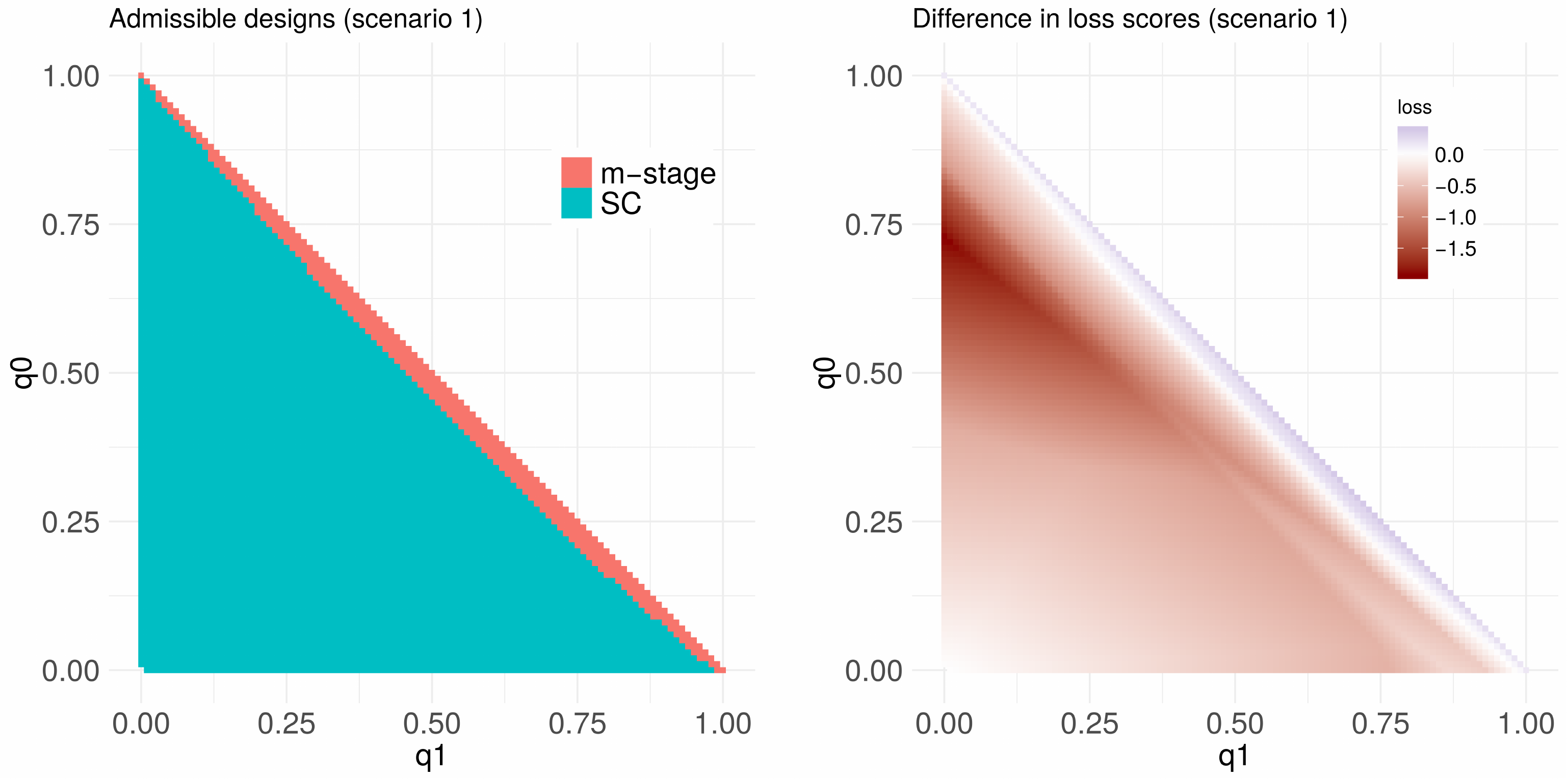}
    \caption{The omni-admissible design and difference in loss scores between the SC and $m$-stage admissible designs (positive favours $m$-stage), scenario 1 $(\alpha, \beta, p_0, p_1)=(0.05, 0.15, 0.10, 0.30)$.}\label{fig:scen1}
\end{figure}

The design with the lowest expected loss across all those compared, the omni-admissible design, is shown in Figure \ref{fig:scen1} (a). Across all combinations of weights, the omni-admissible design is either the SC design or the $m$-stage design. The difference in expected loss between SC and $m$-stage admissible designs is shown in Figure \ref{fig:scen1} (b). It shows that while the $m$-stage design has a considerably lower loss score near the triangle's hypotenuse, that is, when there is low weight on maximum sample size $N$, the SC has a similar loss score to the $m$-stage design otherwise, especially as weight of $N$ approaches the maximum. The maximum difference between a superior SC design and inferior $m$-stage design is 3.0. The range of expected loss across all admissible designs of all design types is (14.1, 79.4), with median 23.1 (IQR [19.2, 26.6]).

\begin{figure}[!h]
    \centering
        \includegraphics[width=0.9\textwidth]{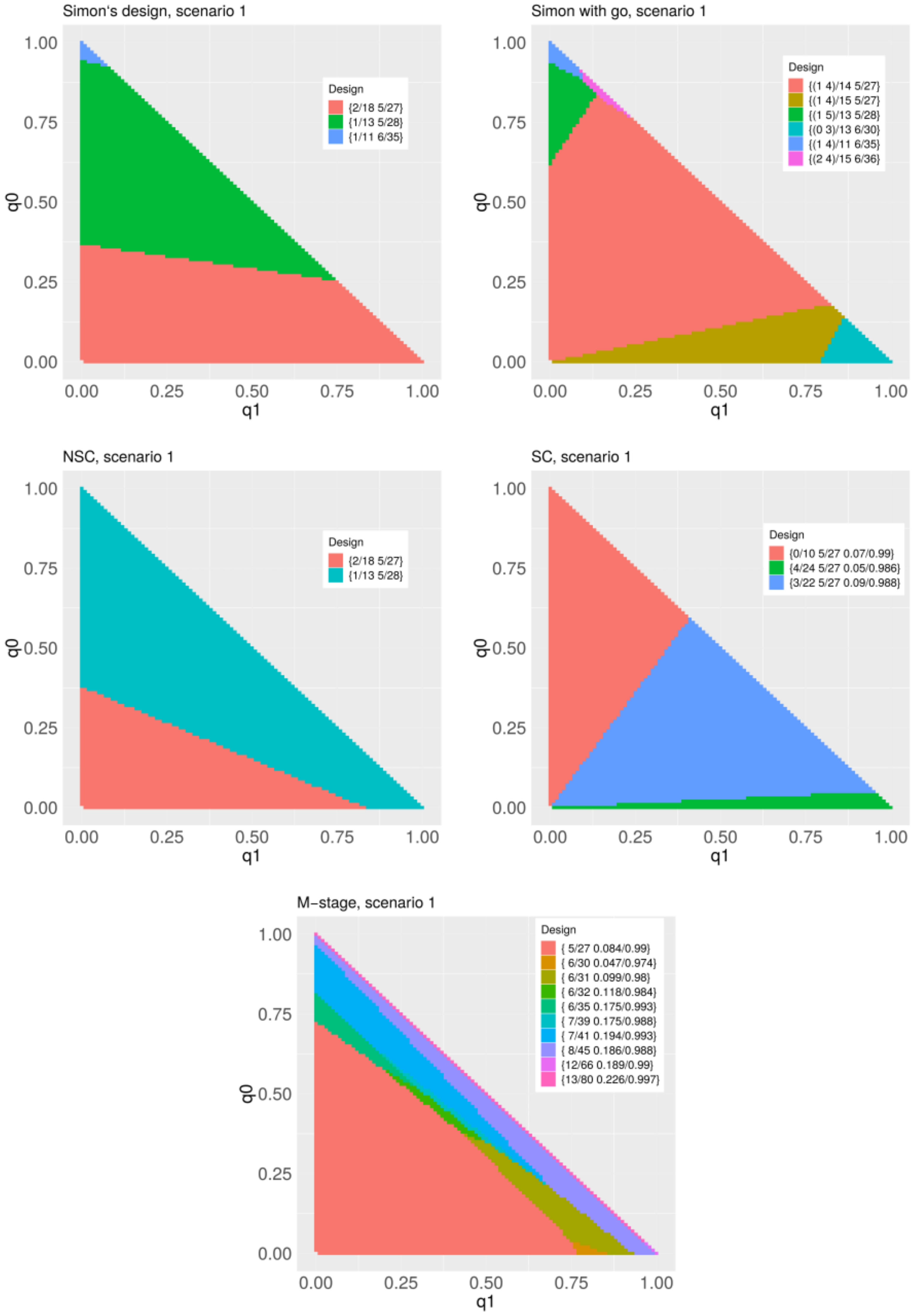}
    \caption{Admissible designs for scenario 1 $(\alpha, \beta, p_0, p_1)=(0.05, 0.15, 0.10, 0.30)$. Designs: (a) Simon's design; (b) Simon with go; (c) NSC; (d) SC; (e) $m$-stage.}\label{fig:design_1}
\end{figure}

\subsection{Admissible designs, by design type --- scenario 1 only}
In Figure \ref{fig:design_1}, the admissible designs are shown for each combination of weights, for scenario 1. For completeness, corresponding figures are shown for scenarios 2 and 3 in the supplementary material. The plots of admissible designs for Simon's and Simon with go designs, Figures \ref{fig:design_1}(a) and \ref{fig:design_1}(b), match those obtained by \cite{mander} as do the corresponding plots in the supplementary material. The overall results are similar across all three scenarios: the proposed designs generally contain a greater number of admissible designs across the combinations of weights examined than existing designs. For the proposed designs, the admissible design regions often contain slopes parallel to the hypotenuse, suggesting that the admissible design may be more dependent on the weight of $N$ than $E(N|p_0)$ or $E(N|p_1)$. In some cases, this is manifested in long, thin regions near the hypotenuse. At the hypotenuse, the admissible designs have the greatest maximum sample size of all the possible admissible designs, with maximum sample size decreasing as the weight of $N$ increases (that is, towards the bottom left corner), as could be expected. When the weight on $N$ is not close to one, the novel designs often have a maximum sample size similar to those that do not employ curtailment.

\subsection{Scenario 2: Design parameters $(\alpha, \beta, p_0, p_1)=(0.05, 0.2, 0.1, 0.3)$}
\begin{table}[ht]
\centering
\begin{tabular}{rcccccccccccc}
  \toprule
 & $r_1$ & $e_1$ & $n_1$ & $r$ & $n_2$ & $n$ & $E(N|p_0)$ & \%S$_{p_0}$ & $E(N|p_1)$ & \%S$_{p_1}$ & $\theta_F$ & $\theta_E$ \\ 
  \midrule $\mathbf{H_0}$\textbf{-optimal} \\ \midrule
Simon & 1 &  & 10 & 5 & 19 & 29 & 15.0 & 1.00 & 26.2 & 1.00 &  &  \\ 
  Simon go & 1 & 4 & 10 & 5 & 19 & 29 & 15.0 & 1.00 & 23.3 & 0.89 &  &  \\ 
  NSC & 1 &  & 10 & 5 & 19 & 29 & 14.1 & 0.94 & 17.1 & 0.66 & 0.000 & 1.000  \\ 
  SC & 5 &  & 33 & 7 & 10 & 43 & 11.7 & 0.78 & 13.3 & 0.51 & 0.216 & 0.994 \\ 
  $m$-stage &  &  &  & 9 &  & 53 & 11.7 & 0.78 & 12.9 & 0.49 & 0.216 & 0.991 \\ 
   \midrule $\mathbf{H_1}$\textbf{-optimal} \\Simon & 2 &  & 18 & 5 & 7 & 25 & 19.9 & 1.00 & 24.6 & 1.00 &  &  \\ 
  Simon go & 0 & 3 & 13 & 5 & 11 & 24 & 20.8 & 1.05 & 17.5 & 0.71 &  &  \\ 
  NSC & 1 &  & 10 & 5 & 19 & 29 & 14.1 & 0.71 & 17.1 & 0.70 & 0.000 & 1.000 \\ 
  SC & 3 &  & 19 & 8 & 24 & 43 & 12.5 & 0.63 & 13.0 & 0.53 & 0.163 & 0.980 \\ 
  $m$-stage &  &  &  & 13 &  & 76 & 11.7 & 0.59 & 12.8 & 0.52 & 0.219 & 0.992 \\ 
   \midrule \\Wald &  &  &  &  &  &  & 11.5 & 0.58 & 12.4 & 0.50 &  &  \\ 
   \midrule $\mathbf{H_0}$\textbf{-minimax} \\Simon & 1 &  & 15 & 5 & 10 & 25 & 19.5 & 1.00 & 24.6 & 1.00 &  &  \\ 
  Simon go & 2 & 4 & 19 & 5 & 5 & 24 & 20.3 & 1.04 & 20.2 & 0.82 &  &  \\ 
  NSC & 1 &  & 15 & 5 & 10 & 25 & 18.4 & 0.94 & 18.4 & 0.75 & 0.000 & 1.000 \\ 
  SC & 0 &  & 9 & 5 & 16 & 25 & 15.3 & 0.79 & 14.6 & 0.59 & 0.058 & 0.973 \\ 
  $m$-stage &  &  &  & 5 &  & 25 & 15.5 & 0.79 & 14.6 & 0.59 & 0.090 & 0.972 \\ 
   \midrule $\mathbf{H_1}$\textbf{-minimax}  \\Simon & 2 &  & 18 & 5 & 7 & 25 & 19.9 & 1.00 & 24.6 & 1.00 &  &  \\ 
  Simon go & 0 & 3 & 13 & 5 & 11 & 24 & 20.8 & 1.05 & 17.5 & 0.71 &  &  \\ 
  NSC & 2 &  & 18 & 5 & 7 & 25 & 18.8 & 0.95 & 18.4 & 0.75 & 0.000 & 1.000 \\ 
  SC & 0 &  & 9 & 5 & 16 & 25 & 15.3 & 0.77 & 14.6 & 0.59 & 0.058 & 0.973 \\ 
  $m$-stage &  &  &  & 5 &  & 25 & 15.5 & 0.78 & 14.6 & 0.60 & 0.090 & 0.972 \\ 
   \bottomrule
\end{tabular}
\caption{Admissible designs for each design type, Scenario 2: $(\alpha, \beta, p_0, p_1)=(0.05, 0.20, 0.10, 0.30)$. For all designs, requisite type I error and power is reached. Columns \%S$_{p_0}$ and \%S$_{p_1}$ show expected sample size as a proportion of Simon's design under $p=p_0$ and $p=p_1$ respectively.} 
\label{tab:scenario2}
\end{table}

Scenario 2 decreases the required power by 0.05 to 0.80. Table \ref{tab:scenario2} shows the optimal designs for each design across the four specified criteria. The two proposed designs outperform the existing designs under $H_0$- and $H_1$-optimal.  Under $H_0$- and $H_1$-minimax, Simon with go has a lower maximum sample size ($n=24 \text{ vs } n=25$ for all others), though $E(N|p_0)$ and $E(N|p_1)$ for the proposed designs are considerably lower. Under all four criteria, the expected sample sizes of the proposed designs under both the null and alternative hypotheses are lower than those of the existing designs, and the thresholds satisfy  $\theta_F < 0.22$ and $\theta_E > 0.97$. Again, the expected sample sizes of Wald are comparable to those of the $m$-stage design, and again the SC design happens to be the $m$-stage design with the addition of one explicit interim analysis.

\begin{figure}[h]
    \centering
        \includegraphics[width=\textwidth]{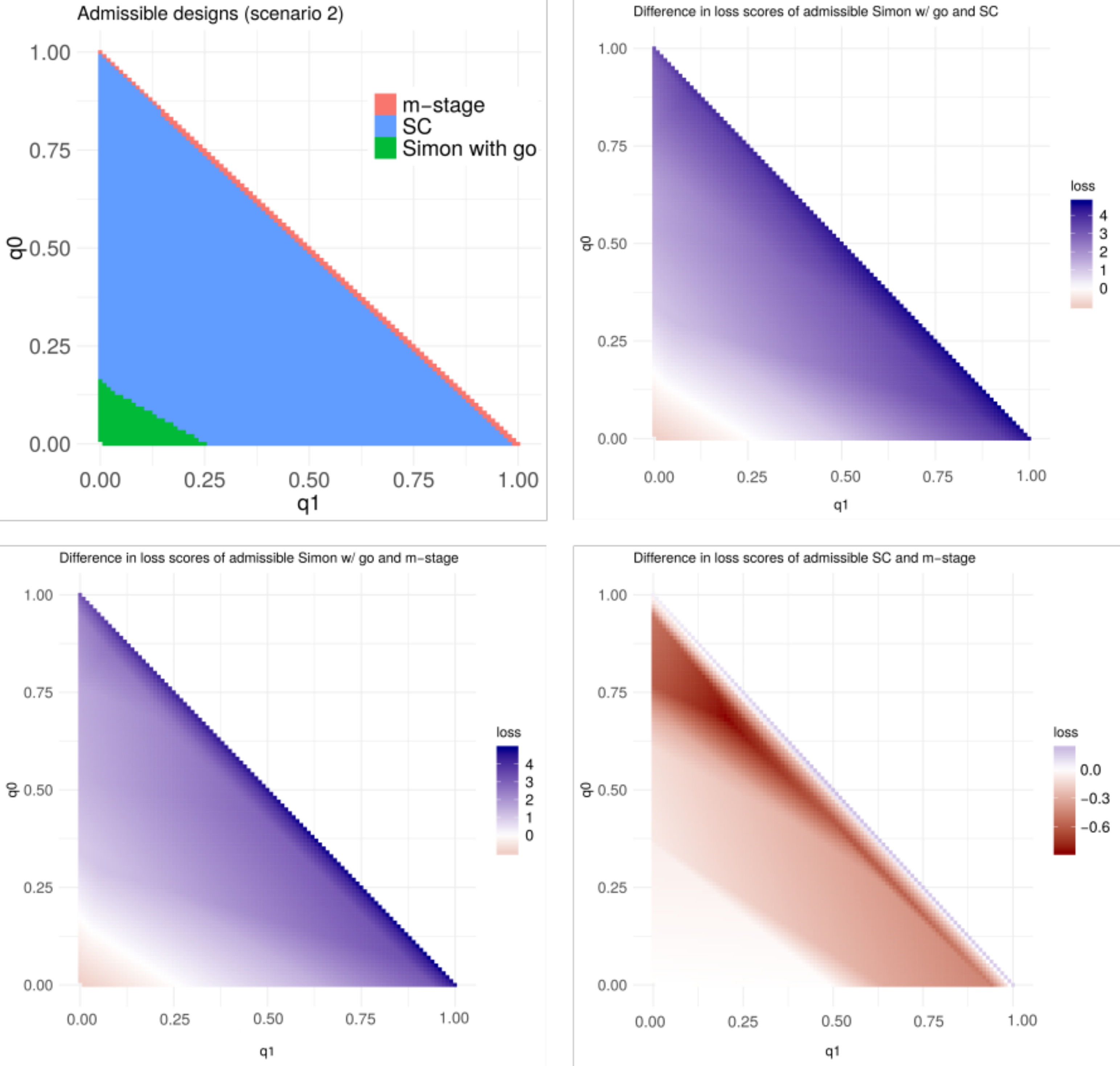}
     \caption{Omni-admissible design and difference in loss scores between the following pairs of admissible designs: Simon with go and SC (positive favours SC), Simon with go and $m$-stage (positive favours $m$-stage) and SC and $m$-stage (positive favours $m$-stage). Scenario 2 $(\alpha, \beta, p_0, p_1)=(0.05, 0.20, 0.10, 0.30)$.}\label{fig:scen2}
\end{figure}

Figure \ref{fig:scen2}(a) shows that the omni-admissible design is SC for most sets of weights, with exceptions where the weight of $N$ is either close to one (as Simon with go has the lowest $N$ of any design) or close to zero, where $m$-stage is superior. Figures \ref{fig:scen2}(b) and \ref{fig:scen2}(c) show that even in the region where the Simon with go design is superior, the loss scores of the SC and $m$-stage admissible designs are similar and as such, should be considered comparable in terms of optimality. The maximum difference in loss score in favour of Simon with go compared to both the SC and $m$-stage designs is 1.0. Figure \ref{fig:scen2}(d) shows that the difference in expected loss between SC and $m$-stage is small, less than 0.9 at all points, while the range of loss scores across all admissible designs in this scenario (11.7, 75.4), with median 20.4 (IQR [17.3, 23.2]).

\subsection{Scenario 3: Design parameters $(\alpha, \beta, p_0, p_1)=(0.05, 0.2, 0.2, 0.4)$}
\begin{table}[ht]
\centering
\begin{tabular}{rcccccccccccc}
  \toprule
 & $r_1$ & $e_1$ & $n_1$ & $r$ & $n_2$ & $n$ & $E(N|p_0)$ & \%S$_{p_0}$ & $E(N|p_1)$ & \%S$_{p_1}$ & $\theta_F$ & $\theta_E$ \\ 
  \midrule $\bm{H_0}$\textbf{-optimal} \\ 
Simon & 3 &  & 13 & 12 & 30 & 43 & 20.6 & 1.00 & 37.9 & 1.00 &  &  \\ 
  Simon go & 3 & 7 & 13 & 12 & 30 & 43 & 20.5 & 1.00 & 35.0 & 0.92 &  &  \\ 
  NSC & 3 &  & 13 & 12 & 30 & 43 & 18.8 & 0.91 & 27.8 & 0.73 & 0.000 & 1.000 \\ 
  SC & 9 &  & 35 & 13 & 12 & 47 & 15.1 & 0.73 & 20.5 & 0.54 & 0.222 & 0.996 \\ 
  $m$-stage &  &  &  & 17 &  & 60 & 15.0 & 0.73 & 18.9 & 0.50 & 0.219 & 0.993 \\ 
   \midrule $\bm{H_1}$\textbf{-optimal} \\Simon & 4 &  & 18 & 10 & 15 & 33 & 22.3 & 1.00 & 31.6 & 1.00 &  &  \\ 
  Simon go & 3 & 6 & 16 & 11 & 19 & 35 & 23.1 & 1.04 & 24.8 & 0.78 &  &  \\ 
  NSC & 4 &  & 18 & 10 & 15 & 33 & 20.4 & 0.92 & 25.1 & 0.80 & 0.000 & 1.000 \\ 
  SC & 13 &  & 44 & 14 & 3 & 47 & 15.8 & 0.71 & 19.1 & 0.60 & 0.146 & 0.986 \\ 
  $m$-stage &  &  &  & 19 &  & 65 & 15.1 & 0.68 & 18.7 & 0.59 & 0.209 & 0.990 \\ 
   \midrule \\Wald &  &  &  &  &  &  & 14.7 & 0.66 & 18.2 & 0.58 &  &  \\ 
   \midrule $\bm{H_0}$\textbf{-minimax} \\Simon & 4 &  & 18 & 10 & 15 & 33 & 22.3 & 1.00 & 31.6 & 1.00 &  &  \\ 
  Simon go & 2 & 6 & 15 & 10 & 17 & 32 & 24.9 & 1.12 & 24.9 & 0.79 &  &  \\ 
  NSC & 4 &  & 18 & 10 & 15 & 33 & 20.4 & 0.92 & 25.1 & 0.80 & 0.000 & 1.000 \\ 
  SC & 0 &  & 11 & 10 & 21 & 32 & 21.3 & 0.96 & 20.9 & 0.66 & 0.050 & 0.985 \\ 
  $m$-stage &  &  &  & 10 &  & 32 & 21.5 & 0.96 & 20.9 & 0.66 & 0.050 & 0.985 \\ 
   \midrule $\bm{H_1}$\textbf{-minimax}  \\Simon & 4 &  & 18 & 10 & 15 & 33 & 22.3 & 1.00 & 31.6 & 1.00 &  &  \\ 
  Simon go & 2 & 6 & 15 & 10 & 17 & 32 & 24.9 & 1.12 & 24.9 & 0.79 &  &  \\ 
  NSC & 4 &  & 18 & 10 & 15 & 33 & 20.4 & 0.92 & 25.1 & 0.80 & 0.000 & 1.000 \\ 
  SC & 0 &  & 11 & 10 & 21 & 32 & 21.3 & 0.96 & 20.9 & 0.66 & 0.050 & 0.985 \\ 
  $m$-stage &  &  &  & 10 &  & 32 & 21.5 & 0.96 & 20.9 & 0.66 & 0.050 & 0.985 \\ 
   \bottomrule
\end{tabular}
\caption{Admissible designs for each design type, Scenario 3: $(\alpha, \beta, p_0, p_1)=(0.05, 0.20, 0.20, 0.40)$. For all designs, requisite type I error and power is reached. Columns \%S$_{p_0}$ and \%S$_{p_1}$ show expected sample size as a proportion of Simon's design under $p=p_0$ and $p=p_1$ respectively.} 
\label{tab:scenario3}
\end{table}

Scenario 3 increases both $p_0$ and $p_1$ by 0.1, resulting in the design parameters $(\alpha, \beta, p_0, p_1) = (0.05, 0.2, 0.2, 0.4)$. Table \ref{tab:scenario3} shows the optimal designs for the four optimality criteria. The proposed designs outperform the existing designs under all four criteria. In particular, the minimum maximum sample sizes of the proposed designs are lower than those of the Simon and NSC designs. The admissible $m$-stage designs under $H_0$- and $H_1$-optimal have comparable expected sample sizes to that of Wald, under both the null and alternative hypotheses. The admissible designs for $H_0$- and $H_1$-minimax are identical. The conditional power thresholds of the proposed designs satisfy $\theta_E > 0.98$ and $\theta_F < 0.23$ for all designs. 

\begin{figure}
    \centering
       \includegraphics[width=\textwidth]{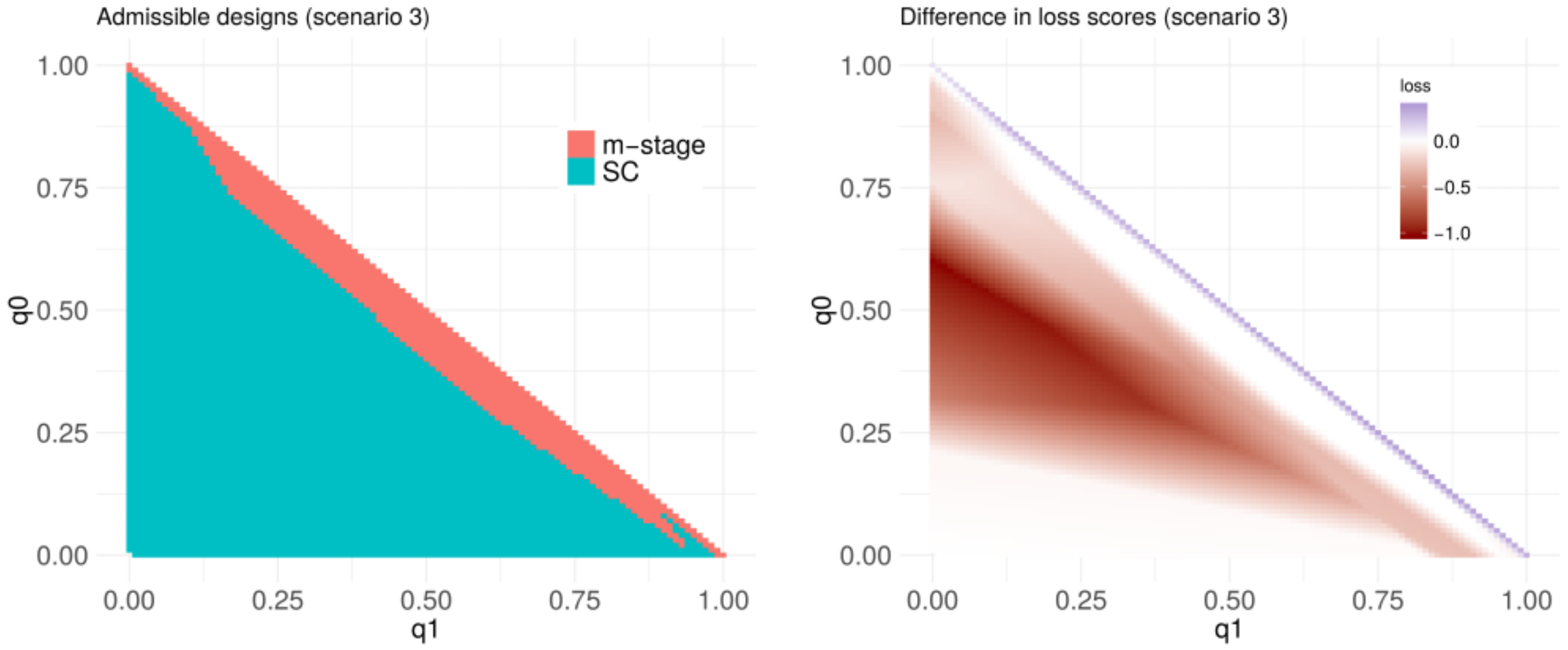}
    \caption{Omni-admissible design, and difference in loss scores between SC and $m$-stage designs (positive favours $m$-stage), scenario 3 $(\alpha, \beta, p_0, p_1)=(0.05, 0.20, 0.20, 0.40)$.}\label{fig:scen3}
\end{figure}

Figure \ref{fig:scen3}(a) shows that each omni-admissible design across the range of possible weights is again either SC or $m$-stage. The difference in expected loss between these two designs is shown in Figure \ref{fig:scen3}(b). The difference is small, $<1.1$ at all points, compared to the range of loss scores across all admissible designs in this scenario (15.0, 64.5), with median 26.4 (IQR [22.7, 30.4]).

\begin{figure}[h]
    \centering
    \includegraphics[width=\textwidth, center]{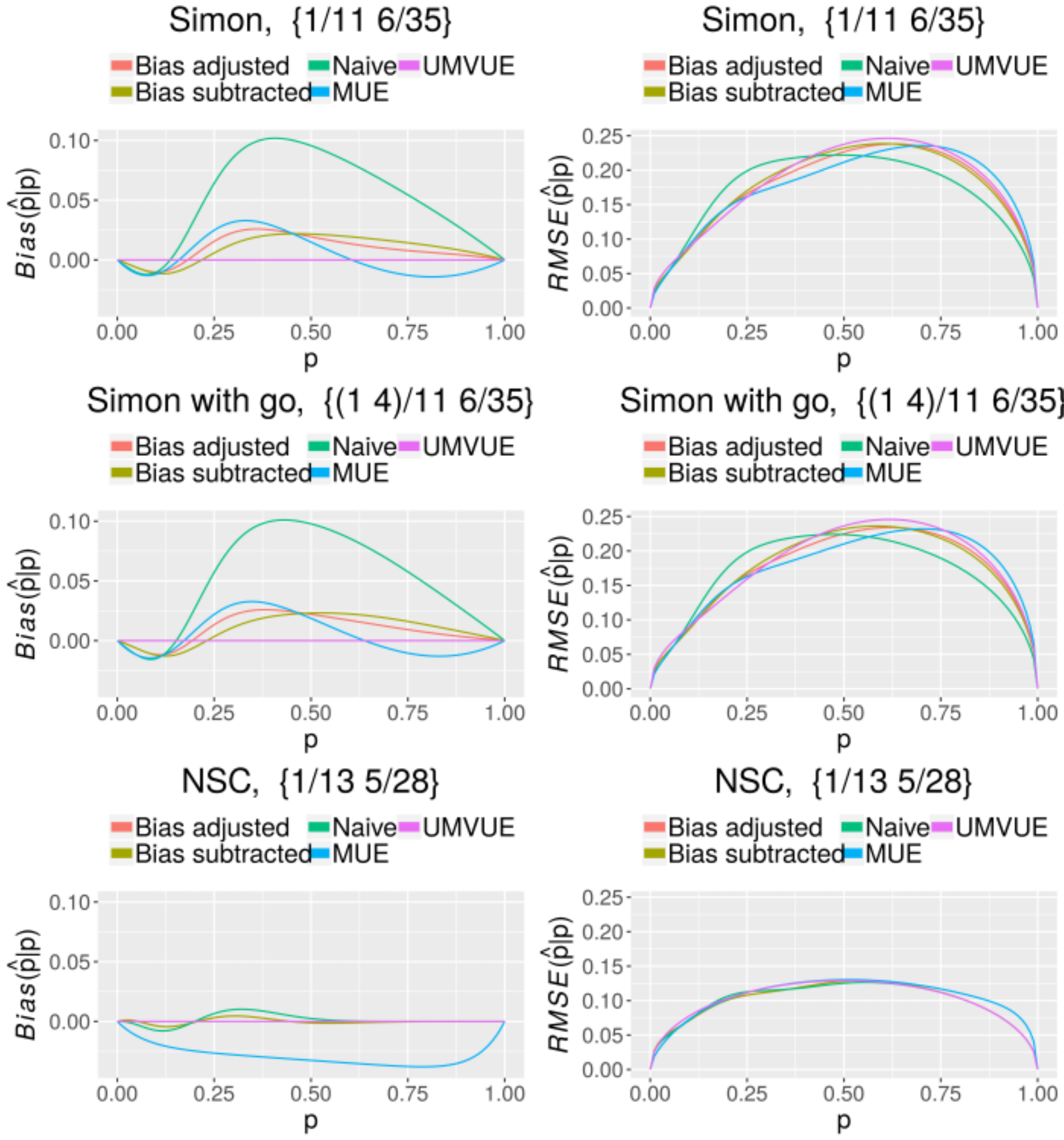}
    \caption{Bias, RMSE for $H_0$-optimal designs, scenario 1 $(\alpha, \beta, p_0, p_1)=(0.05, 0.15, 0.10, 0.30)$. Designs: (a, b) Simon's design; (c, d) Simon with go; (e, f) NSC.}
    \label{fig:bias_rmse_selected}
\end{figure}

\begin{figure}[h]\ContinuedFloat
    \centering
    \includegraphics[width=\textwidth]{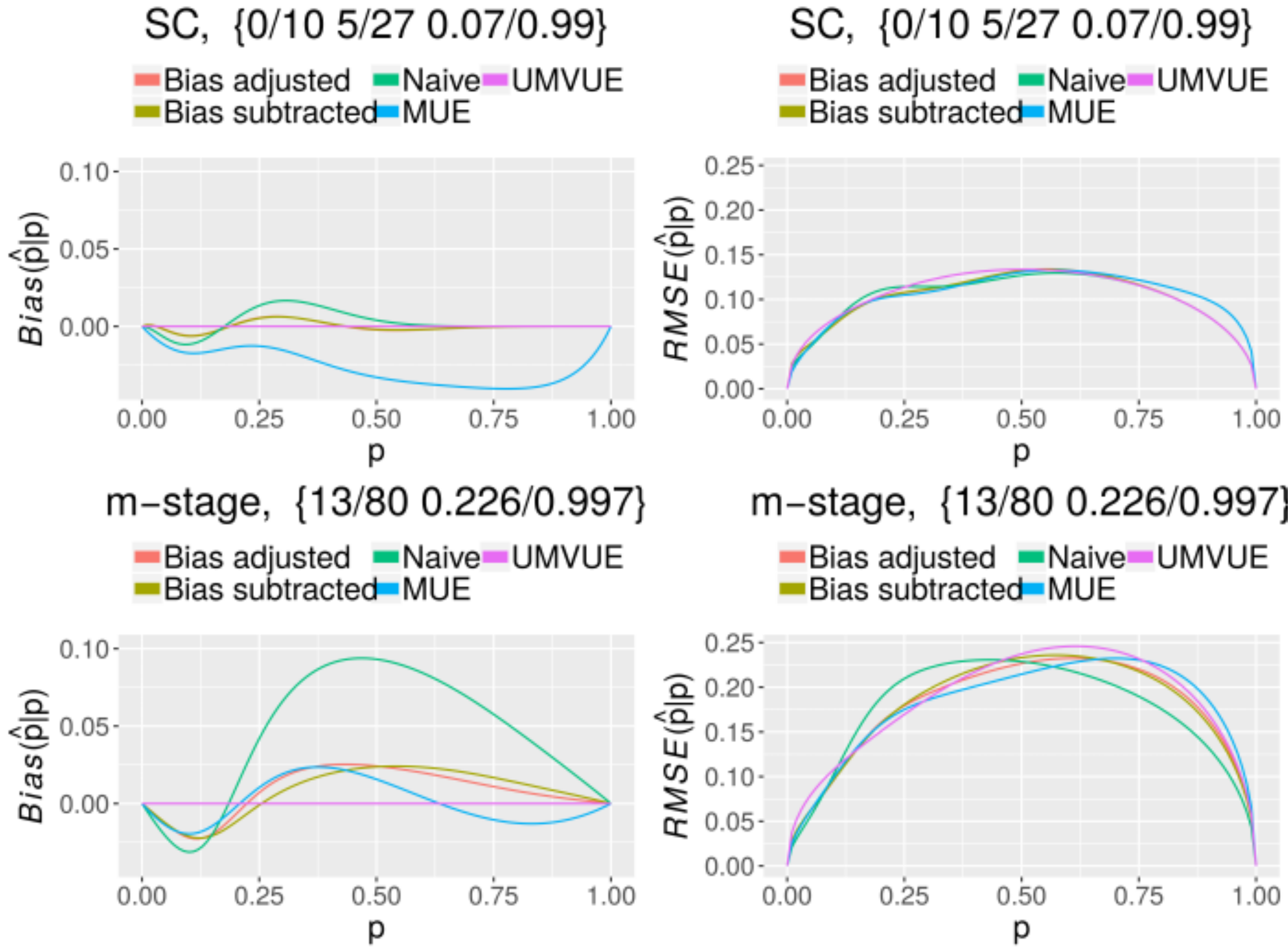}
    \caption{Bias, RMSE for $H_0$-optimal designs, scenario 1 $(\alpha, \beta, p_0, p_1)=(0.05, 0.15, 0.10, 0.30)$, continued. Designs: (g, h) SC; (i, j) $m$-stage.}
    \label{fig:bias_rmse_selected_cont}
\end{figure}    

\subsection{Continuous monitoring in practice and effect of reduced monitoring frequency}

There may be some practical objection to the continuous monitoring required by the SC and $m$-stage designs. However, trial recruitment rate is often lower than expected: in a review of 122 trials, \cite{Campbell2007} found that early participant recruitment was slower than expected in 77 (63\%) of 122 reviewed trials, and a review of 151 randomised controlled trials by \cite{Walters2017} reported a median recruitment rate of 0.92 participants per centre per month. Further, \cite{Campbell2007} found that only 38 (31\%) of 122 trials reached their intended sample size and 66 (54\%) requested a trial extension. As such, trial recruitment rates are generally lower than anticipated, and low enough to facilitate continuous monitoring, especially as no additional statistical analysis is required to use stochastic curtailment; all stopping boundaries are obtained in advance. 

If trial recruitment rate is anticipated to be so great as to make continuous monitoring impractical, a design permitting stochastic curtailment only after every block of $B$ participants may be considered. Such an approach can still produce savings in expected sample size. As an example, Table \ref{tab:block} shows admissible designs for blocks of size four and eight, that is, permitting curtailment after every four or eight participants respectively, for the first scenario $(\alpha, \beta, p_0, p_1)=(0.05, 0.15, 0.10, 0.30)$. These are shown alongside the admissible designs for the Simon design and the $m$-stage design. The $m$-stage design may be considered as using blocks of size one. Under the $H_0$- and $H_1$-optimal criteria, the designs permitting stochastic curtailment only after every four or eight participants produce considerable savings in expected sample size compared to using Simon's design. Under the $H_0$- and $H_1$-minimax criteria, which are combined in the table as the admissible designs are identical in this instance, savings in expected sample size are again made, with the exception of $E(N|p_0)$ when using block size eight.

\begin{table}[ht]
\centering
\begin{tabular}{rccccccccccc}
  \toprule
 & $r_1$ &  $n_1$ & $r$ & $n_2$ & $n$ & $E(N|p_0)$ & \%S$_{p_0}$ & $E(N|p_1)$ & \%S$_{p_1}$ & $\theta_F$ & $\theta_E$ \\ 
  \midrule $\bm{H_0}$\textbf{-optimal} \\ 
  Simon & 1 &   11 & 6 & 24 & 35 & 18.3 & 1.00 & 32.3 & 1.00 &  &  \\ 
  $m$-stage &    &  & 13 &  & 80 & 14.1 & 0.77 & 14.4 & 0.45 & 0.226 & 0.997 \\
  Block size 4  & & & 10 &  & 56 & 14.5 & 0.79 & 16.3 & 0.50 & 0.534 & 0.988 \\
  Block size 8  & & & 12 &  & 72 & 16.1 & 0.88 & 19.4 & 0.60 & 0.691 & 0.991 \\
   \midrule $\bm{H_1}$\textbf{-optimal} \\
  Simon 	& 2   & 18 & 5 & 9 & 27 & 20.4 & 1.00 & 26.5 & 1.00 &  &  \\ 
  $m$-stage &    &  & 12 &  & 66 & 14.3 & 0.70 & 14.4 & 0.54 & 0.189 & 0.990 \\ 
  Block size 4  & & & 11 &  & 64 & 14.7 & 0.72 & 16.1 & 0.61 & 0.550 & 0.991 \\
  Block size 8  & & & 16 &  & 80 & 16.8 & 0.82 & 18.2 & 0.69 & 0.559 & 0.974 \\   \midrule $\bm{H_{0/1}}$\textbf{-minimax} \\
   Simon & 2   & 18 & 5 & 9 & 27 & 20.4 & 1.00 & 26.5 & 1.00 &  &  \\ 
  $m$-stage   &  &  & 5 &  & 27 & 18.7 & 0.92 & 16.6 & 0.63 & 0.084 & 0.990 \\ 
  Block size 4  & & & 6 &  & 32 & 18.8 & 0.92 & 18.7 & 0.71 & 0.194 & 0.984 \\
  Block size 8  & & & 6 &  & 32 & 21.3 & 1.04 & 21.7 & 0.82 & 0.340 & 0.988 \\
   \bottomrule
\end{tabular}
\caption{Selection of admissible designs, including stochastically curtailed designs with stopping permitted after every four and eight patients, for scenario 1: $(\alpha, \beta, p_0, p_1)=(0.05, 0.15, 0.10, 0.30)$. For all designs, requisite type I error and power is reached. Columns \%S$_{p_0}$ and \%S$_{p_1}$ show expected sample size as a proportion of Simon's design under $p=p_0$ and $p=p_1$ respectively.} 
\label{tab:block}
\end{table}

\subsection{Estimation (scenario 1, selected)}

Bias and root mean squared error (RMSE) in the response rate estimates are shown in Figure \ref{fig:bias_rmse_selected}, with the maximum absolute bias and RMSE shown in Table \ref{tab:maxbias}. For brevity, the only plots shown are for the $H_0$-optimal designs for scenario 1. In Simon's and the Simon with go designs, the estimates of bias are close to zero for all estimators (Figures \ref{fig:bias_rmse_selected}(a), \ref{fig:bias_rmse_selected}(c)). For designs that employ curtailment, the bias adjusted, bias subtracted MUE and UMVUE estimators remain close to zero, while the na\"ive estimator gives less accurate estimates (Figures \ref{fig:bias_rmse_selected}(e), \ref{fig:bias_rmse_selected_cont}(g), \ref{fig:bias_rmse_selected_cont}(i)). Overall, bias and RMSE is only slightly poorer among the proposed designs than the Simon-based designs for $p < p_1$. For greater $p$, the poorer estimates among the proposed designs are a result of the trial being curtailed with fewer patients compared to the Simon-based designs. The maximum absolute bias is similar across designs, with the exception of somewhat greater bias among the two SC designs under the na\"ive estimator (Table \ref{tab:maxbias}).

The RMSE of the estimates gradually increase as the degree of early stopping increases, with the increase being more pronounced at less likely response rates, though RMSE decreases sharply to zero when the response rate is close to one. The maximum RMSE for the two SC designs is somewhat greater than the rest (Table \ref{tab:maxbias}).

\begin{table}[ht]
\small
\centering
\begin{tabular}{rrrrrr|rrrrr}
  & \multicolumn{5}{c}{Bias (absolute)} & \multicolumn{5}{c}{RMSE} \\ \toprule
 & Bias adj. & Bias subt. & Na\"ive & MUE & UMVUE & Bias adj. & Bias subt. & Na\"ive & MUE & UMVUE \\ 
  \midrule
Simon & 0.01 & 0.01 & 0.03 & 0.03 & 0.00 & 0.10 & 0.10 & 0.10 & 0.11 & 0.10 \\ 
  Simon$_{GO}$ & 0.01 & 0.01 & 0.03 & 0.05 & 0.00 & 0.15 & 0.15 & 0.14 & 0.15 & 0.15 \\ 
  NSC & 0.01 & 0.01 & 0.04 & 0.03 & 0.00 & 0.16 & 0.17 & 0.16 & 0.16 & 0.17 \\ 
  SC & 0.02 & 0.03 & 0.09 & 0.03 & 0.00 & 0.23 & 0.24 & 0.22 & 0.22 & 0.24 \\ 
  $m$-stage & 0.03 & 0.02 & 0.09 & 0.02 & 0.00 & 0.23 & 0.24 & 0.23 & 0.23 & 0.25 \\ 
   \bottomrule
\end{tabular}
\caption{Maximum absolute bias and RMSE for various point estimators of $H_0$-optimal designs, scenario 1: $(\alpha, \beta, p_0, p_1)=(0.05, 0.15, 0.1, 0.3)$.} 
\label{tab:maxbias}
\end{table}

\section{Discussion}

This paper introduces two new designs for binary outcome, single-arm phase II clinical trials, one based on Simon's design and one based on the single-stage design. These designs propose allowing early stopping to make a go or no-go decision before the final decision would otherwise be certain, known as stochastic curtailment.

In addition to these proposed designs, this paper also introduces four approaches for improving a search for an optimal design: firstly, the exact distribution of the trial outcomes is obtained, allowing the trial's operating characteristics to be known without simulation error. Secondly, a new approach is proposed for finding relevant stopping boundaries when using stochastic curtailment, based on the conditional power at each point in each possible trial, allowing more potential designs to be evaluated. Thirdly, the conditional power at each point in each potential design is calculated taking the possibility of stochastic curtailment into account; it is not calculated based on an approximation that does not account for stopping due to stochastic curtailment. Finally, type I error and power are only calculated after taking curtailment into account; no designs are discarded in advance for not reaching the required type I error and power in their uncurtailed form. Between them, these four concepts serve dual purposes: to allow more potential designs to be examined without excessive computational intensity, and to increase the accuracy of the reported operating characteristics of such designs. This should result in investigators being able to make a more efficient choice of design for any potential study.

The proposed designs, combined with the above approaches, were compared to a number of existing curtailed designs. They were compared in a real data example, where they were shown to approximately halve the trial sample size from 19 to between 8 and 11, and also across three scenarios with regard to the following optimality criteria: $H_0$-optimal, $H_1$-optimal, $H_0$-minimax and $H_1$-minimax. With the exception of the $H_0/H_1$-minimax criteria in one scenario, where the proposed designs had a maximum sample size of 25 compared to 24 in an existing design, the proposed designs were superior across all criteria and scenarios. For the proposed designs, the expected sample size under the $H_0$-optimal and $H_1$-optimal criteria was comparable to those obtained using Wald's sequential probability ratio test, with a difference of less than a single participant in general, in favour of Wald's test. However, while this test may result in favourable expected sample size, a design with no maximum sample size would be impractical for clinical trials, where a maximum sample size is necessary due to limited resources, population size and so on. 	

The proposed designs were also compared to existing designs across a combination of multiple criteria, using a weighted loss function. Employing Mander et al's expected loss function, admissible designs were obtained for each design over a grid of combination of weights for the criteria maximum sample size, expected sample size under the null hypothesis and expected sample size under the alternative hypothesis. For each possible combination of weights, the design realisation that had the lowest expected loss across all designs was recorded, and the design to which it belonged was recorded. This design has been termed the omni-admissible design. 
Plotting the omni-admissible design for each possible combination of weights, it is shown that the proposed designs are almost always better in terms of expected loss. While the two-stage stochastic curtailment design can be superior to the $m$-stage design, the difference is generally slight. However, we accept that increasing the maximum $N$ searched over is likely to find design realisations with lower expected sample size, and the disparity between the SC and $m$-stage searches in terms of maximum $N$ searched over may be the reason why the omni-admissible design is an $m$-stage rather than an SC design for some combinations of weights. Searching for $m$-stage admissible designs is approximately two orders of magnitude faster than the SC design, with a full search able to be conducted in under 60 minutes for $N \in [20, 80]$. For this reason, we recommend that investigators focus on this design. The effect of reducing the frequency of monitoring, was examined. It was shown that considerable savings in expected sample size may still be made even when employing designs with less frequent monitoring.


There may be some apprehension regarding ending a trial before the final decision is certain compared to a different design. However, the trials are powered taking this into account, in the same way that Simon's design meets the required type I error and power despite allowing the potential for stopping before the final decision is certain compared to a single-stage trial. Indeed, we have shown that a Simon design may end for a no-go decision even when the probability of success for an effective treatment in a single-stage design is as much as 0.80. 
 Further, across all admissible designs obtained from the proposed designs, the threshold for stopping for a go decision is close to one, $\theta_E > 0.97$. Finally, if necessary, it is possible to allow the specification of limits to the thresholds $\theta_F$ and $\theta_E$ to ranges that are acceptable to the investigator.
 
These design types could be generalised in a number of ways in future work. In case of a desire to collect a certain degree of information in a trial, a trial could be specified to end only after data is available for a minimum of $n$ participants. Secondly, estimates of confidence intervals and p values could be compared to those from existing design types. Finally, the effect of multi-participant cohorts could be examined.

In summary, this paper proposes two designs for phase II, single arm, binary outcome clinical trials, and augments them with a number of approaches for finding better designs, using exact distributions so that the designs' operating characteristics can be obtained without simulation error. These designs, combined with the proposed approaches, have been shown to be superior to existing designs, both when considering a single optimality criterion and when considering a weighted combination of multiple criteria.

The data that support the findings of this study are available from the corresponding author upon reasonable request.



  \bibliographystyle{elsarticle-harv} 


\bibliography{law_bib}

%
%
%

\end{document}


\maketitle

\section*{Choosing thresholds}

For a single-stage NSC design, the number of possible CP values (including zero and one) is given by

\begin{equation*}
(r+1)(N-r)+1,
\end{equation*}

which is a quadratic equation that reaches a maximum at $r=(N-1)/2$. The number of possible CPs increases linearly with $N$. 
A'Hern \cite{ahern} states without proof that the final stopping boundary for a single-stage trial with no curtailment will be approximately 

\begin{equation}\label{eq:ahern}
r = N(p_0 + [(z_\alpha / (z_\alpha + z_{1-\beta})) \times (p_1 - p_0)]).
\end{equation}

For a single-stage trial, $N=40$ and design parameters $(\alpha, \beta, p_0, p_1)=(0.05, 0.20, 0.10, 0.30)$, the approximate stopping boundary given by Equation (\ref{eq:ahern}) is $r=9.5967$. Setting $r$ equal to the smallest integer greater than this, 10, such a trial would have 331 possible CP values, resulting in 54,615 ordered pairs $(\theta_F, \theta_E)$ such that $\theta_F < \theta_E$.  

For a two-stage NSC design, the number of possible CP values is given by

\begin{equation*}
(r_1 + 1)(n_1 - r_1) + (r-r_1)(N-r) - 1.
\end{equation*}



It is possible to employ constraints that further reduce the the number of possible ordered pairs, using the following argument: If a trial using SC reaches its penultimate stage, i.e. $m=N-1$, without a decision being made, then the trial will result in a go decision if the final participant responds and a no-go decision if the final participant does not respond. 
The conditional power at this point, $CP(p_1)_{r, N-1}$, is equal to $p_1$. Under SC, a trial stops for a no-go decision if $CP < \theta_F$. However, if the true response rate is great enough to warrant further study, then the probability of a go decision at this point is $p \geq p_1$, and so the trial should never be curtailed for a no-go decision at this point. As such, $\theta_F$ should be constrained such that $\theta_F < p_1$. Introducing this constraint reduces the the number of possible ordered pairs $(\theta_F, \theta_E)$  in the above two-stage example to 12,084. Finally, for all admissible designs incorporating SC across all scenarios, $\theta_E > 0.97$. With this in mind, constraining the search to require  $\theta_E > 0.95$ reduces the number of ordered pairs to 7,809. This is comparable to the number of ordered pairs that would be produced when searching over the uniform sequence \{0, 0.01, ... 1\}, which is 5,050.



\section*{Admissible design search}


The computational intensity of searching for admissible designs increases with the number of possible trials, so it is of interest to use sensible constraints for the range of $r$.
A'Hern states without proof that for a single-stage design without curtailment, the trial end stopping boundary $r$ lies in the interval $\left[ Np_0, Np_1 \right]$ \cite{ahern}; as such, the final stopping boundary for the proposed designs is constrained to this interval. As an alternative, the boundaries for Wald's sequential probability ratio test may be used as constraints \cite{wald}; This is a design with no maximum sample size $N$: the trial simply continues until a go or no-go decision can be made with the requisite level of certainty. Wald derives lower and upper stopping boundaries for the sequential probability ratio test to be

\begin{equation*}
S_{NO GO} = \frac{1}{D}\left(\text{log }\frac{\beta}{1 - \alpha} + m \text{ log }\frac{1-p_0}{1-p_1} \right)
\end{equation*}

\begin{equation*}
S_{GO} = \frac{1}{D}\left(\text{log }\frac{1-\beta}{\alpha} + m \text{ log }\frac{1-p_0}{1-p_1} \right),
\end{equation*}

where

\begin{equation*}
D = \text{log }\frac{p_1}{p_0} - \text{log } \frac{1-p_1}{1-p_0}
\end{equation*}

after $m$ participants. All admissible designs using the proposed designs all three scenarios were obtained using A'Hern's boundaries, and were also within the more strict boundaries of Wald. We therefore recommend that Wald's boundaries are used to constrain $r$, to reduce computational intensity.

\section*{Wald's sequential probability ratio test}
Wald's sequential probability ratio test, introduced directly above, results in a lower expected sample size than many existing design types with the same choice of type I error and power \cite{wald}. 
As such, it is worthwhile to compare how close the expected sample size of a given design is to the expected sample size obtained using this test. Hence for each scenario, the expected sample sizes of the $H_0$- and $H_1$-optimal admissible designs were compared to those of Wald.

For this design, the expected sample size under $p=p_0$ is

\begin{equation*}
E(N|p_0) = \frac{(1-\alpha)\text{ log } \frac{\beta}{1-\alpha} + \alpha \text{ log } \frac{1-\beta}{\alpha}}{p_0 \text{ log }\frac{p_1}{p_0} + (1-p_0) \text{ log } \frac{1-p_1}{1-p_0}}.
\end{equation*}

The expected sample size under $p=p_1$ is 

\begin{equation*}
E(N|p_1) = \frac{\beta \text{ log } \frac{\beta}{1-\alpha} + 1-\beta \text{ log } \frac{1-\beta}{\alpha}}{p_1 \text{ log }\frac{p_1}{p_0} + (1-p_1) \text{ log } \frac{1-p_1}{1-p_0}}.
\end{equation*}


\section*{Designs that are unfeasible in their uncurtailed form}

The type I error and power of uncurtailed designs are not calculated prior to adding curtailment, as a design that is feasible after SC may not be feasible before the incorporation of SC. For example, take the design parameters $(\alpha, \beta, p_0, p_1) = (0.05, 0.2, 0.1, 0.4)$. The single-stage design $(N, r)=(21, 4)$ for $(p_0, p_1)=(0.1, 0.4)$ has operating characteristics $(\alpha, \beta)=(0.052, 0.037)$ (rounded to three d.p.), which is not feasible as $\alpha>0.05$. However, applying SC by using the thresholds $(\theta_F, \theta_E)=(0.31744, 0.9919024)$ results in the operating characteristics $(\alpha, \beta)=(0.048, 0.141)$ (rounded to three d.p.), which is feasible and has $E(N|p_0)=7.5, E(N|p_1)=7.6$ (rounded to two d.p).

\section*{Expected loss by design}

\begin{figure}[h]
    \centering
        \includegraphics[width=\textwidth]{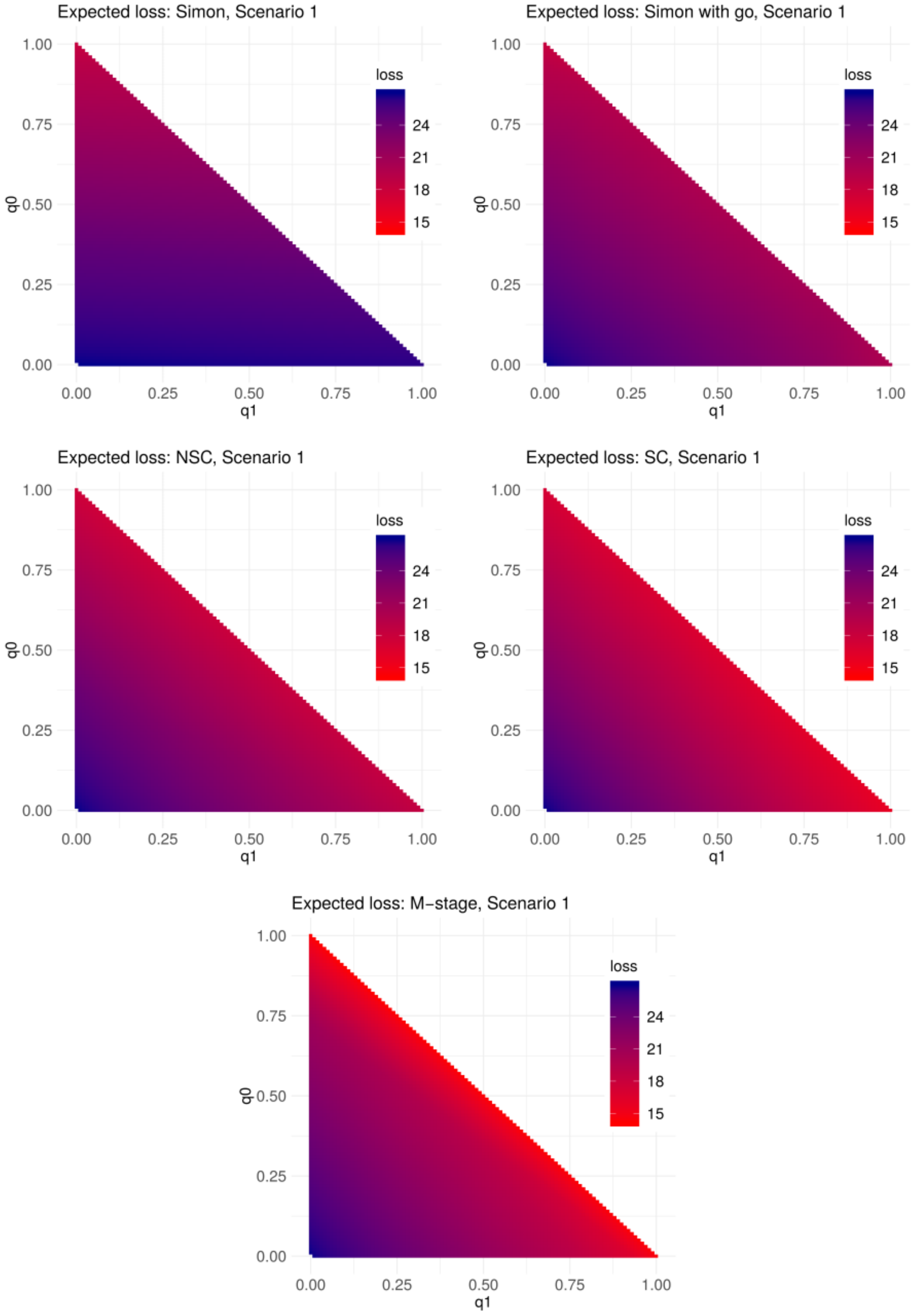}
    \caption{Expected loss for each design type: (a) Simon's design; (b) Simon with go; (c) NSC; (d) SC; (e) $m$-stage, for scenario 1 $(\alpha, \beta, p_0, p_1)=(0.05, 0.15, 0.10, 0.30)$.}\label{fig:expected_loss_1}
\end{figure}

\begin{figure}[h]
    \centering
        \includegraphics[width=\textwidth]{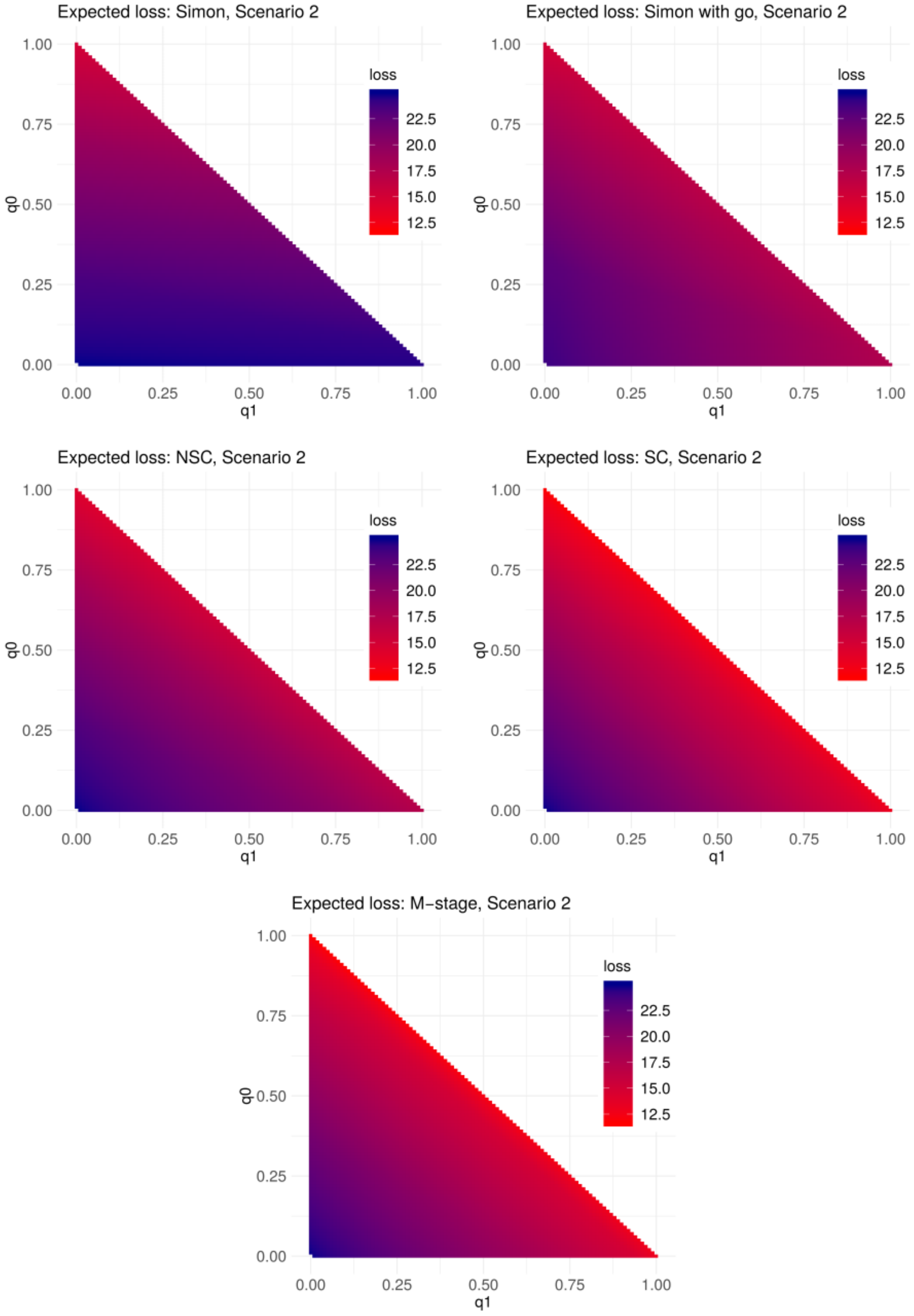}
    \caption{Expected loss for each design type: (a) Simon's design; (b) Simon with go; (c) NSC; (d) SC; (e) $m$-stage, for scenario 2 $(\alpha, \beta, p_0, p_1)=(0.05, 0.20, 0.10, 0.30)$.}\label{fig:expected_loss_2}
\end{figure}

\begin{figure}[h]
    \centering
        \includegraphics[width=\textwidth]{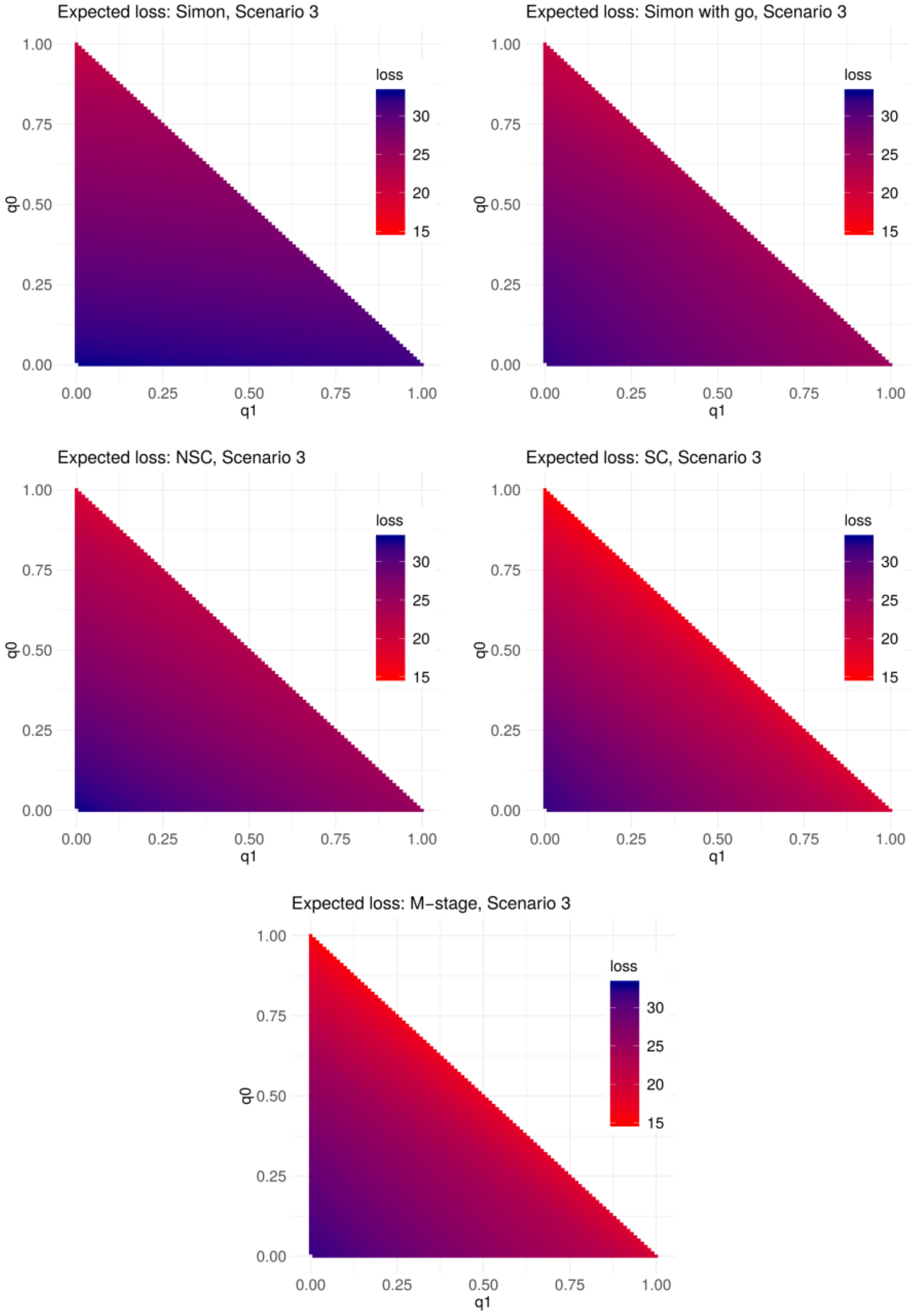}
    \caption{Expected loss for each design type: (a) Simon's design; (b) Simon with go; (c) NSC; (d) SC; (e) $m$-stage, for scenario 3 $(\alpha, \beta, p_0, p_1)=(0.05, 0.20, 0.20, 0.40)$.}\label{fig:expected_loss_3}
\end{figure}

Heat maps of expected loss for the admissible designs of each design are shown in Figures \ref{fig:expected_loss_1}, \ref{fig:expected_loss_2} and \ref{fig:expected_loss_3} for scenarios 1, 2 and 3 respectively. In each scenario, it can be seen that the novel designs, particularly those that employ SC (SC and $m$-stage) have a lower expected loss in general. These design types seem most superior in regions where $N$ is weighted close to 0, (that is, near the hypotenuse) and $E(N|p_0)$ is close to 0 (that is, where $q_0 \approx 0$).

\section*{Admissible designs, by design type (scenarios 2 and 3)}

\begin{figure}[h]
    \centering
        \includegraphics[width=\textwidth]{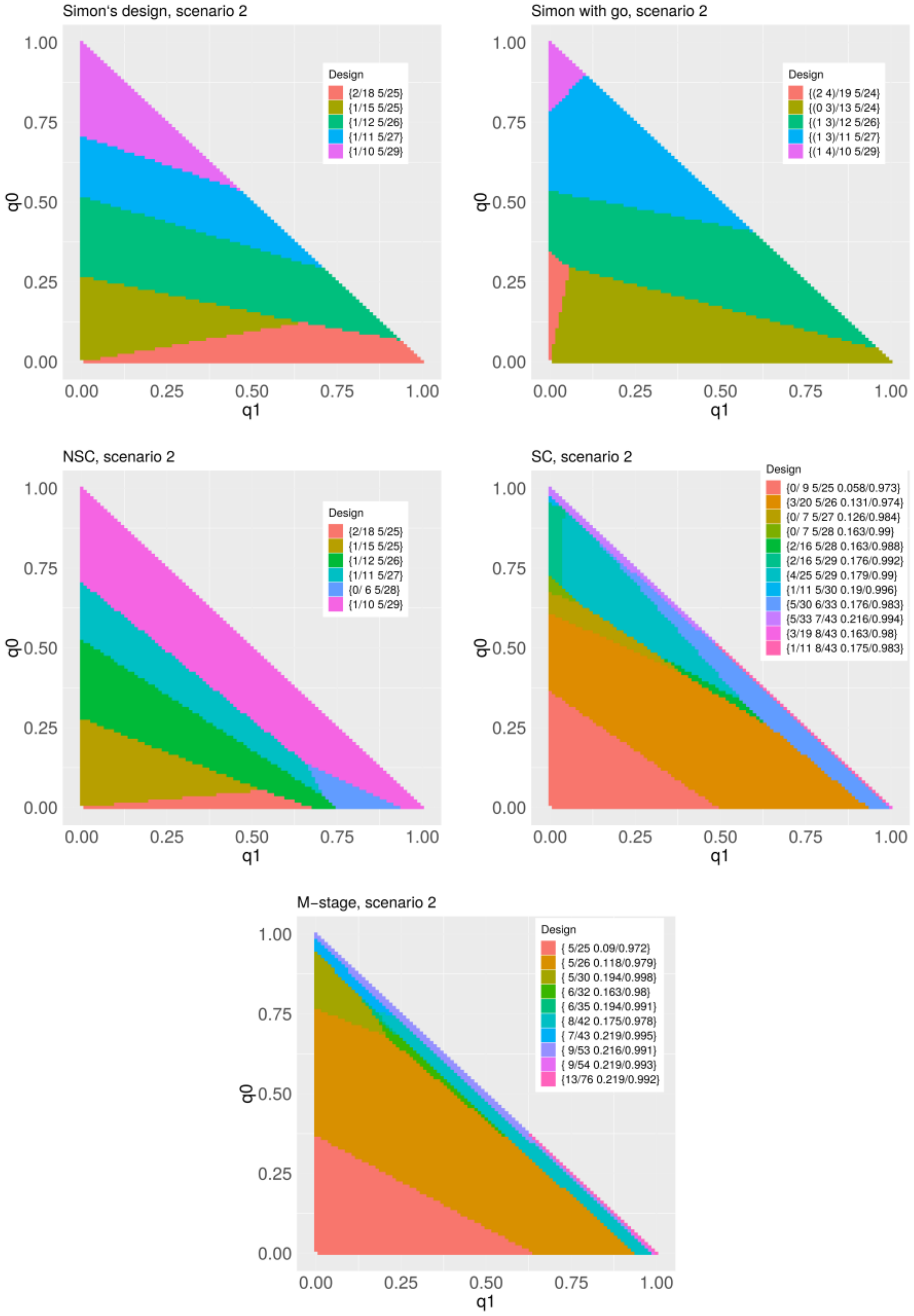}
    \caption{Admissible designs for each design type: (a) Simon's design; (b) Simon with go; (c) NSC; (d) SC; (e) $m$-stage, for scenario 2 $(\alpha, \beta, p_0, p_1)=(0.05, 0.20, 0.10, 0.30)$.}\label{fig:design_2}
\end{figure}

\begin{figure}[h]
    \centering
        \includegraphics[width=\textwidth]{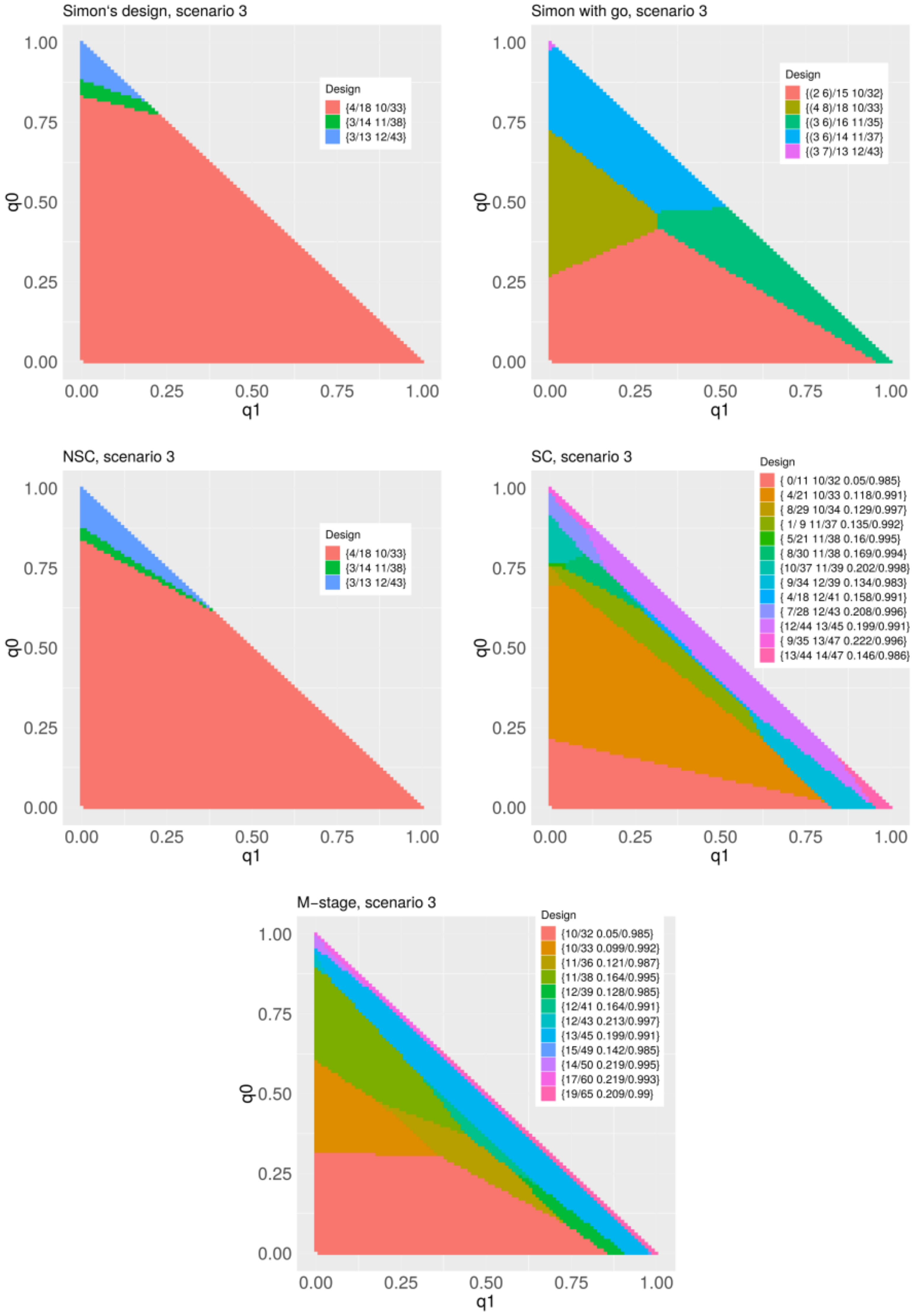}
    \caption{Admissible designs for each design type: (a) Simon's design; (b) Simon with go; (c) NSC; (d) SC; (e) $m$-stage, for scenario 3 $(\alpha, \beta, p_0, p_1)=(0.05, 0.20, 0.20, 0.40)$.}\label{fig:design_3}
\end{figure}

For completeness, the range of admissible designs for each compared design for scenarios 2 and 3 is shown in Figures \ref{fig:design_2} and \ref{fig:design_3}.

\section*{Point estimators for multi-stage trials}

The RMSE can be obtained using the bias and the variance of the estimate of the response rate. With the exception of the UMVUE, the adjusted estimators for response rate have not been described for multi-stage trials. The bias-subtracted and bias-adjusted estimators are described in terms of the expected value of the response rate and its bias. The expected estimate of the response rate, $\hat{p}$, can be obtained by taking the product of the observed response rate for each possible terminal point and its probability given some true $p$, and summing across all possible terminal points:

\begin{equation*}
E(\hat{p}|p, \mathbf{r}, \mathbf{a}, \mathbf{n}) = \displaystyle\sum\limits_{g=1}^{m} \displaystyle\sum\limits_{S=0}^{n_g} \hat{p}(S_{n_g}, n_g)U(S_{n_g}, n_g | p, \mathbf{r}, \mathbf{a}, \mathbf{n}),
\end{equation*}

where $\mathbf{r}=r_1, r_2, \dots, r_m$ and $\mathbf{a}=a_1, a_2, \dots, a_m$ are the vectors of stopping boundaries for go and no-go decisions respectively at each stage, and $\mathbf{n}=n_1, n_2, \dots, n_m$ is the vector of sample size at each stage. The bias, variance and RMSE are as follows:
\begin{align*}
\text{Bias}(\hat{p}|p, \mathbf{r}, \mathbf{a}, \mathbf{n}) & = E(\hat{p}|p, \mathbf{r}, \mathbf{a}, \mathbf{n}) - p \\
\text{Var}(\hat{p}|p, \mathbf{r}, \mathbf{a}, \mathbf{n}) & = E(\hat{p}^2|p, \mathbf{r}, \mathbf{a}, \mathbf{n}) - E(\hat{p}|p, \mathbf{r}, \mathbf{a}, \mathbf{n})^2 \\
\text{RMSE}(\hat{p}|p, \mathbf{r}, \mathbf{a}, \mathbf{n}) & = \sqrt{\text{Bias}(\hat{p}|p, \mathbf{r}, \mathbf{a}, \mathbf{n})^2 + \text{Var}(\hat{p}|p, \mathbf{r}, \mathbf{a}, \mathbf{n})}
\end{align*}

As stated in the main body of the paper, the na\"ive estimator for $p$ is simply $\hat{p}_{naive}=S_m/m$. The bias-subtracted estimator is then
\begin{equation*}
p_{bias-sub} = \hat{p}_{naive} - \text{Bias}(\hat{p}_{naive}|\hat{p}_{naive}, \mathbf{r}, \mathbf{a}, \mathbf{n})
\end{equation*}

The bias-adjusted estimator is the numerical solution to 
\begin{equation*}
	p_{bias-adj} = \hat{p}_{naive} - \text{Bias}(\hat{p}_{naive}|\hat{p}_{bias-adj}, \mathbf{r}, \mathbf{a}, \mathbf{n})
\end{equation*}

Finally, the median unbiased estimator, $\hat{p}_{MUE}$ is obtained by numerically searching for the value of $p$ that would make the p-value equal to 0.5:

\begin{equation*}
\text{p-val}(S_m, m|\hat{p}_{MUE}) = 0.5,
\end{equation*}

where the p-value is computed as the sum of the probability of possible outcomes with a larger value of the UMVUE. The estimates for all estimators are obtained using the R package \textit{singlearm} (https://github.com/mjg211/singlearm).